\documentclass{aa}  

\usepackage{graphicx}
\usepackage{txfonts}
\usepackage{xcolor}
\usepackage[symbol]{footmisc}
\usepackage{geometry}
\usepackage{natbib}
\usepackage{amssymb}

\usepackage{arydshln} 

\begin{document} 

   \title{Long period modulation of the classical T Tauri star CI Tau:} 

   \subtitle{evidence for an eccentric close-in massive planet at 0.17\,au}

   \author{R. Manick \inst{1}
          \and
           A. P. Sousa \inst{1}
          \and 
          J. Bouvier\inst{1}
           \and 
           J.M. Almenara \inst{1,4}
          \and
          L. Rebull \inst{2}
          \and
          A. Bayo \inst{5}
           \and
          A. Carmona \inst{1}
          \and
          E. Martioli \inst{3, 11}
          \and
         L. Venuti \inst{7}
          \and
          G. Pantolmos \inst{1}
          \and 
           \'A. K\'osp\'al \inst{6,9,10}
           \and
          C. Zanni \inst{8}
          \and
          X. Bonfils \inst{1}
          \and 
          C. Moutou \inst{2}
          \and
          X. Delfosse \inst{1} 
          \and
          the SLS consortium \thanks{Based on observations performed with CFHT at Maunakea, Hawaii, ExTrA facility at La Silla Observatory in Chile, K2 mission and LCOGT.}
          }

   \institute{$^{1}$ Univ. Grenoble Alpes, CNRS, IPAG, 38000 Grenoble, France\\
              $^{2}$ Infrared Science Archive (IRSA), IPAC, 1200 E. California Blvd., California Institute of Technology, Pasadena, CA 91125, USA \\
              $^{3}$ Laborat\'orio Nacional de Astrofis\'ica, Rua Estados Unidos 154, 37504-364 Itajub\'a, MG, Brazil \\
               $^{4}$ Observatoire de Gen\`eve, D\'epartement d’Astronomie, Universit\'e de Gen\`eve, Chemin Pegasi 51b, 1290 Versoix, Switzerland\\
               $^{5}$ European Organisation for Astronomical Research in the Southern Hemisphere (ESO), Karl-Schwarzschild-Str. 2, 85748 Garching bei München, Germany\\
               $^{6}$ Konkoly Observatory, Research Centre for Astronomy and Earth Sciences, Eötvös Loránd Research Network (ELKH), Konkoly-Thege Miklós út 15-17, 1121, Budapest, Hungary\\
               $^{7}$ SETI Institute, 339 Bernardo Ave, Suite 200, Mountain View, CA 94043, USA \\
               $^{8}$ INAF – Osservatorio Astrofisico di Torino, Strada Osservatorio 20, 10025 Pino Torinese, Italy \\
               $^{9}$ ELTE E\"otv\"os Lor\'and University, Institute of Physics, P\'azm\'any P\'eter s\'et\'any 1/A, 1117 Budapest, Hungary \\
               $^{10}$ Max Planck Institute for Astronomy, K\"onigstuhl 17, 69117 Heidelberg, Germany\\
               $^{11}$ Institut d'Astrophysique de Paris, CNRS, UMR 7095, Sorbonne Universit'{e}, 98 bis bd Arago, 75014 Paris, France \\
              \email{rajeev.manick@univ-grenoble-alpes.fr}\\
             }

   \date{}

 
  \abstract
  {Detecting planets within protoplanetary disks around young stars is essential for understanding planet formation and evolution. However, planet detection using the radial velocity method faces challenges due to strong stellar activity in these early stages.}
   {We aim to detect long-term periodicities in photometric and spectroscopic time series of the classical T Tauri star (CTTS) CI Tau, and retrieve evidence for inner embedded planets in its disk.}
 {The study conducted photometric and spectroscopic analyses using K2 and Las Cumbres Observatory Global Network light curves, and high-resolution spectra from ESPaDOnS and SPIRou. We focus our radial velocity analysis on a wavelength domain less affected by spot activity. To account for spot effects, a quasi-periodic Gaussian process model was applied to K2 light curve, ESPaDOnS, and SPIRou radial velocity data. Additionally, a detailed bisector analysis on cross-correlation functions was carried out to understand the cause of long-term periodicity.}
   {We detect coherent periods at $\sim$\,6.6\,d, 9\,d, $\sim$\,11.5\,d, $\sim$\,14.2\,d and $\sim$\,25.2\,d, the latter is seen consistently across all datasets. Bisector analysis of the cross-correlation functions provides strong hints for combined activity-induced and Doppler reflex signal in the radial velocities at a period of 25.2\,d. Our analysis suggests that this periodicity is best explained by the presence of a 3.6 $\pm$ 0.3 M$_{\rm Jup}$, eccentric (e$\sim$0.58) planet at a semi-major axis of 0.17 au.}
   {We report the detection of a massive inner-planet in CI Tau. Our study outlines the difficulty of searching for disk-embedded planets in the inner 0.1\,au's of young and active systems. We demonstrate that, when searching for planets in actively accreting stars such as CI Tau, the primary limitation is stellar activity rather than the precision of RV measurements provided by the instrument.}


   \maketitle
%

\section{Introduction}
Understanding the processes that govern planet formation and evolution is one of the key questions in planetary science. To address this, several surveys aiming at detecting planets through photometric transits (e.g. \textit{Kepler} and TESS collaborations) have been operational. Most planet discoveries thus far have predominantly centered around main sequence stars \citep{pepe2004}, with a smaller but still significant proportion found around post-main-sequence giants \citep{veras2016}, brown dwarfs \citep{chauvin2004}, intermediate age stars \citep{quinn2014}, and pulsars \citep{wolszczan1992}. Interestingly, only a few dozens of planets have been discovered around younger ($\lesssim$ 10\,Myr) stellar systems, most of them being through direct imaging \citep[e.g.][]{lafreniere2008,hammond2023,wagner2023}, a few being through the transit method \citep[e.g.][]{rowe2014,mann2016} and only one by the radial velocity method \citep[V830 Tau,][]{donati2016}. Another method leading to planet detection around young T Tauri stars (TTSs) was made possible by imaging gaps in their disks \citep[e.g.][]{keppler2018,pinte2019,wang2020}. This method, is limited to detecting them at relatively large distances from their host stars. This limitation underscores the current constraints of the latest generation telescopes, which have yet to achieve the necessary high resolutions to detect planets within the innermost active regions ($\lesssim$\,0.1 au). Consequently, it remains uncertain when and where close-in planets form, as well as the pace at which they migrate either inward or outward through the disk.

The relative paucity of confirmed planetary companions to very young stars highlights the difficulty in detecting them around such stars, particularly in the T Tauri class, which are among the most variable amidst young stars. TTSs are divided into two classes: weak line TTSs (WTTSs) and classical TTSs (CTTSs). WTTSs are systems that lack spectroscopic accretion signatures and generally display rather stable sinusoidal light curves induced by surface features \citep[e.g.][]{stassun1999,grankin2008,frasca2009,donati2014}. Detecting planets using the radial velocity (RV) method in such stable systems is comparatively less challenging than in CTTSs, that display complex characteristics in their light curves (and RVs) as a result of intense stellar activity and/or active accretion phenomena \citep[e.g.][]{vrba1993, herbst1994, rucinski2008, siwak2010, cody2010, alencar2010,fischer2023}.

CI Tau remains the only CTTS in which a disk-embedded close-in planet with $\sim$\,11 M$_{\rm Jup}$ and a period of $\sim$\,9 d was claimed to be detected by \citet{johnskrull2016} and later by \citet{biddle2018}. However, such claims were challenged following Zeeman-Doppler Imaging (ZDI) analysis of ESPaDOnS spectra which demonstrated that the 9\,d modulation corresponds to the stellar rotation period and is induced by stellar activity rather than a purported planet \citep{donati2020}. 

CI Tau is estimated to be approximately 2\,Myr old \citep{guilloteau2014}. \citet{simon2019} estimated a mass of 0.9\,$\pm$\,0.02\,M$_\odot$ based on pre-main sequence (PMS) evolutionary tracks, while an earlier study by \citet{guilloteau2014} reported a slightly lower mass of 0.8\,$\pm$\,0.02\,M$_\odot$, obtained using rotation of its Keplerian disk. The central star has an effective temperature of 4250\,$\pm$\,50\,K \citep{donati2020} and is positioned at 160.3\,$\pm$\,0.4\,pc away from the Sun in the Taurus molecular cloud \citep{vallenari2022}. CI Tau is recognized for hosting a robust, predominantly radial magnetic field of up to 3.7\,kG and displays a varying mass accretion rate of around 2\,$\times$\,10$^{-8}$\,M$_\odot$\,yr$^{-1}$ \citep{donati2020}. On the large scale, the central star is surrounded by a circumstellar disk observable in mm continuum images, extending up to 200\,au, inclined at $\sim$\,55\,$\pm$\,5$^{\circ}$ as reported by \citet{guilloteau2011} and \citet{clarke2018}. This disk exhibits a sequence of dusty rings, with discernible gaps at radii approximately 13, 39, and 100\,au, suggesting ongoing planet formation processes \citep{clarke2018}. On the smaller scale, CI Tau is surrounded by an inner-disk with an inclination of $\sim$\,70$^{\circ}$  \citep{soulain2023}, mis-aligned with the outer disk.

In this paper, we report the detection of a significant periodic signal at $\sim$\,25.2\,d in both photometric (K2 and LCOGT) and spectroscopic time series (ESPaDOnS and SPIRou RVs). The periodic signal in the RVs becomes clearer when stellar activity due to spot modulation is mitigated. In Section \ref{section:observations}, we describe the data sets (photometric and spectroscopic) used in this study. Sections \ref{section:photometric_analysis} and \ref{section:rv_analysis}, details our photometric and spectroscopic time series data analysis, respectively. We discuss our results in Section \ref{section:discussion} and conclude in Section \ref{section:conclusion}.



\section{Observations} \label{section:observations}
\subsection{K2 photometry}
We used publicly available data from the K2 mission \citep[campaign 13,][]{howell2014}. Long cadence Simple Aperture Photometry (SAP) flux light curves was downloaded using the \textsc{lightkurve} package \citep{lightkurve2018} to convert the raw data into a FITS file, from which the usable data was extracted. The full light curve extending over a period of $\sim$\,80\,d, ranging from 8 March 2017 to 27 May 27 2017 UTC (JD 2,457,820 $-$ JD 2,457,901) is shown in the top panel of Figutre \ref{fig:k2lc}.

\subsection{LCOGT multiband photometry} 
Additional photometric time series data for CI Tau was obtained at Las Cumbres Observatory Global Network \citep[LCOGT,][]{brown2013} within 2 epochs separated by two years apart. The first epoch (20B) covered from October 30, 2020 to January 31, 2021 (JD 2,459,153 – JD 2,459,245) where 539 images were acquired in the Sloan g’r’i’ filters over 3 months with a sub-day sampling rate (runs LCO2020B-019, P.I. L. Rebull;  CLN2020B-005, P.I. A. Bayo, for a total of 11 hours). The second epoch (22B/23A) extended from September 27, 2022 to March 31, 2023 (JD 2,459,850 – JD 2,460,033) during which 564 images were acquired over 6 months in the Sloan g'r'i' filters, with a sampling of typically one measurement per day (runs DDT2022B-003, DDT2023A-001; PI L. Rebull, for a total of 11 hours). At both epochs, the g'r'i' images were obtained with the 0.4m SBIG telescopes offering a field-of-view of 29.2$\times$19.5 arcmin, with exposure times of 60, 30, and 30 seconds, respectively. We retrieved the BANZAI reduced images from the LCOGT archive service and the non-calibrated photometric catalogs provided in the image headers for all detected stars in the field. 

In order to compute differential photometry, we considered three stars, 2MASS J04334286+2255065 (JH\,104), 2MASS J04334414+2256179 (JH\,105), and 2MASS J04340717+2251227, located within 6 arcmin of CI Tau and of similar brightness. We used JH\,105 as a comparison star to compute the differential light curve (CI Tau – JH 105), while 2MASS J04340717+2251227 served as a check star to assess these are non-variable sources. The differential light curve (CI Tau – JH 105) was then calibrated in each filter by adopting JH 105’s magnitudes as provided by the APASS survey, namely, g'=13.235, r'=12.238, i’=11.666\,mag. The measurement uncertainty was calculated as rms/sqrt(2) of the differential light curves of the non-variable comparison and check stars. For both epochs, the measurement uncertainty amounts to 0.020, 0.024, and 0.030 mag, in the g’r’i’ filters, respectively. Some measurements of lower quality, due to non-photometric conditions and/or the proximity of the bright Moon were discarded. Specifically, we removed images where the comparison and check stars’ differential photometry indicated a measurement discrepant by more than 2$\sigma$ from the mean. Eventually, for the first epoch, we retained 156 measurements in the g’-band, and 164 in each of the r’ and i’-bands, thus sampling the light curves at a sub-day rate nearly continuously over 93 days during semester 20B. For the second epoch, we gathered 165 measurements in the g’-band, 172 in r’-band, and 175 in i’-band, thus sampling CI Tau’s photometric variability nearly continuously at a daily rate over 183 days during semesters 22B/23A. 

\subsection{ESPaDOnS data}

Spectroscopic observations were conducted using the ESPaDOnS spectropolarimeter at the 3.6\,m Canada-France-Hawaii Telescope (CFHT) on Mauna Kea, Hawaii  (PI: J. F. Donati). A total of 72 spectra were obtained over a time span of $\sim$ 2 months, from mid-December 2016 to mid-February 2017. These observations were divided into 18 sequences, with each sequence comprising 4 consecutive nightly sub-exposures. Each spectrum covers a wavelength range of 3,700 - 10,000 {\AA} at a resolving power of 65,000 \citep{donati2003}. The data were reduced using the standard ESPaDOnS reduction package (Libre-ESpRIT), which is a newer version of the ESpRIT pipeline \citep{donati1997}.

A least-squares deconvolution (LSD) was used to combine a large
number of photospheric spectral lines into a mean LSD profile with an improved S/N \citep{donati1997, kochukhov2010}. The line mask
used for calculating the LSD profiles was extracted from the
VALD database \citep{piskunov1995,kupka1999} for
wavelengths between $\sim$ 3,900 {\AA} and 10,500 {\AA} using a template mask derived
from a synthetic spectrum corresponding to the effective temperature and gravity of CI Tau \citep{guilloteau2014}.

\subsection{SPIRou data} \label{section:spirou}
SPIRou is a fibre-fed, cross-dispersed \'echelle spectropolarimeter mounted on the Canada-France-Hawaii telescope (CFHT) on Maunakea,
Hawaii. The spectrograph provides spectral coverage over a wavelength range from 950 to 2350 nm at a spectral resolving power of 75,000 \citep{donati2020b}.

In total, 425 spectra (4 sub-exposures per night) were obtained for CI Tau during 5 semesters ranging from 18 December 2018 to 13 January 2023 (PI: Jean-Fran\c{c}ois Donati), covering a total time span of $\sim$ 1487 days, but with significant gaps in the RV time series and a varying number of RV points among semesters. We measured the mean signal-to-noise ratio (S/N) in three regions covering wavelengths centred at $\sim$\,1209\,nm, $\sim$\,1557\,nm and $\sim$\,2329\,nm, respectively, to be 40, 76 and 115. We note an overall improvement in S/N with MJD, possibly a result of a combination of the instrument optimization after MJD\,$\sim$\,58900 (optical transmission and guiding precision, C. Moutou, private communication) and an overall increase of the total integration time for the four polarimetric sub-exposures per night, ranging from $\sim$\,1580\,s for the earlier observations (around 2018-2020) to $\sim$\,2000\,s for the newer dates (around 2021-2023). Nevertheless, the spectra taken at earlier dates are also valuable.

\subsection{ESO ExTrA: Low-resolution near-infrared spectroscopy}
We obtained additional data as part of the Exoplanets in Transits and their Atmospheres survey (ExTrA, \citealt{bonfils2015}). The ExTrA facility, situated at La Silla Observatory in Chile, comprises three 60 cm telescopes and a single near-infrared (0.88 to 1.55~$\mu$m) fibre-fed spectrograph.

CI\,Tau was observed on 136 nights between September 28, 2022 and March 21, 2023, using one telescope. Fibre units are located at the focal plane of each telescope, each consisting of two 8\arcsec\ aperture fibers. One fibre is used to observe a star and the other is used to observe the nearby sky background. We used the higher-resolution mode of the spectrograph (R\,$\sim$\,200) and 300-second exposures. We obtained between 2 and 51 exposures per night for a total of 1807 spectra with a median signal-to-noise ratio of 85 at 1.05~$\mu$m. The ExTrA data were corrected for dark current, extracted using the flat-field, corrected for sky background emission, and were wavelength calibrated using custom data reduction software. Median spectra of CI~Tau were computed for each night and telescope, yielding a total of 136 spectra with a median signal-to-noise ratio of 129 and a standard deviation of 40.

\section{Photometric Analysis} \label{section:photometric_analysis}
\subsection{K2 lightcurve} \label{section:GLS_periodogram}

\begin{figure}
   \centering
   \includegraphics[width=9.5cm]{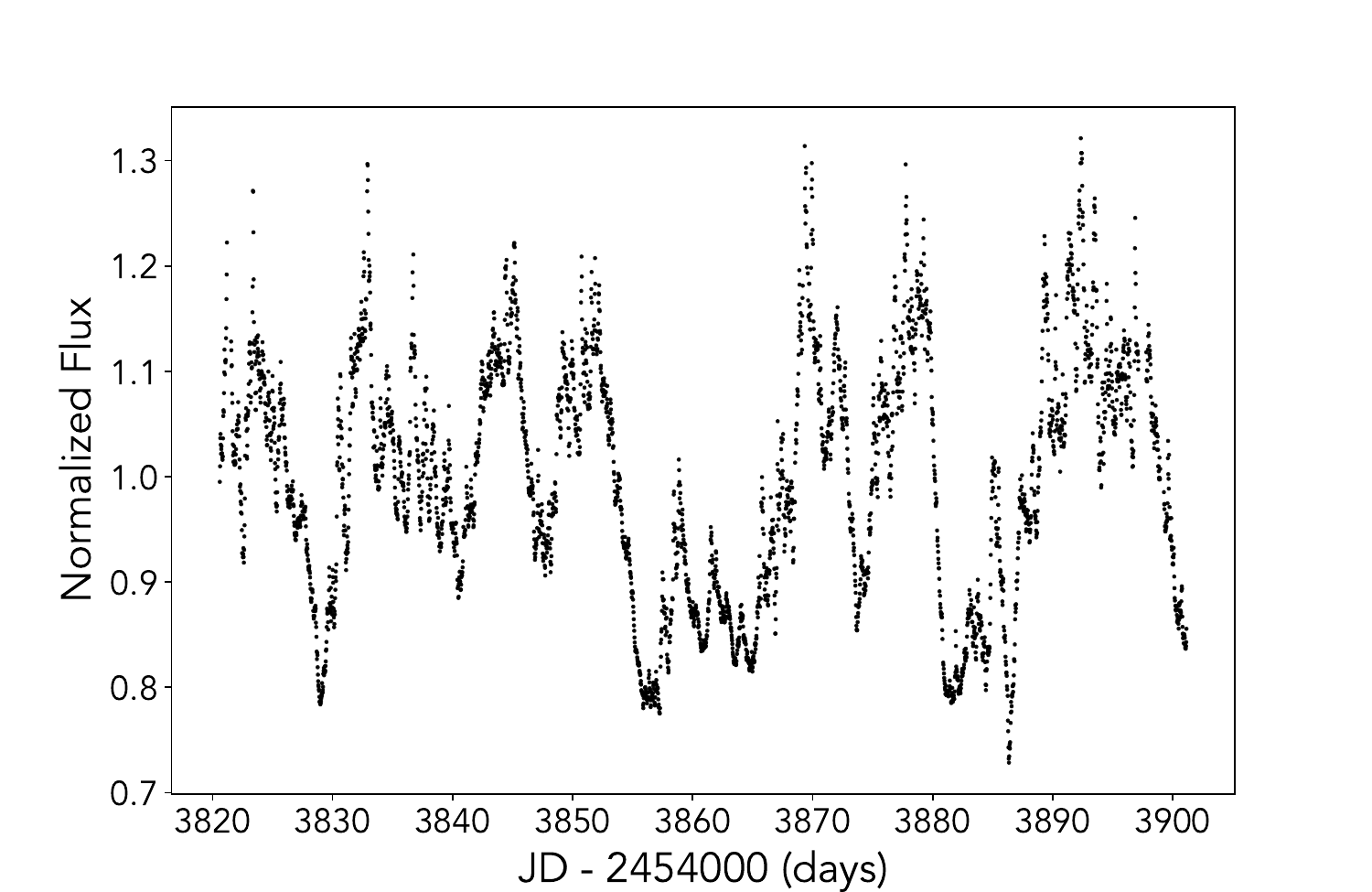} \\
   \includegraphics[width=9.5cm]{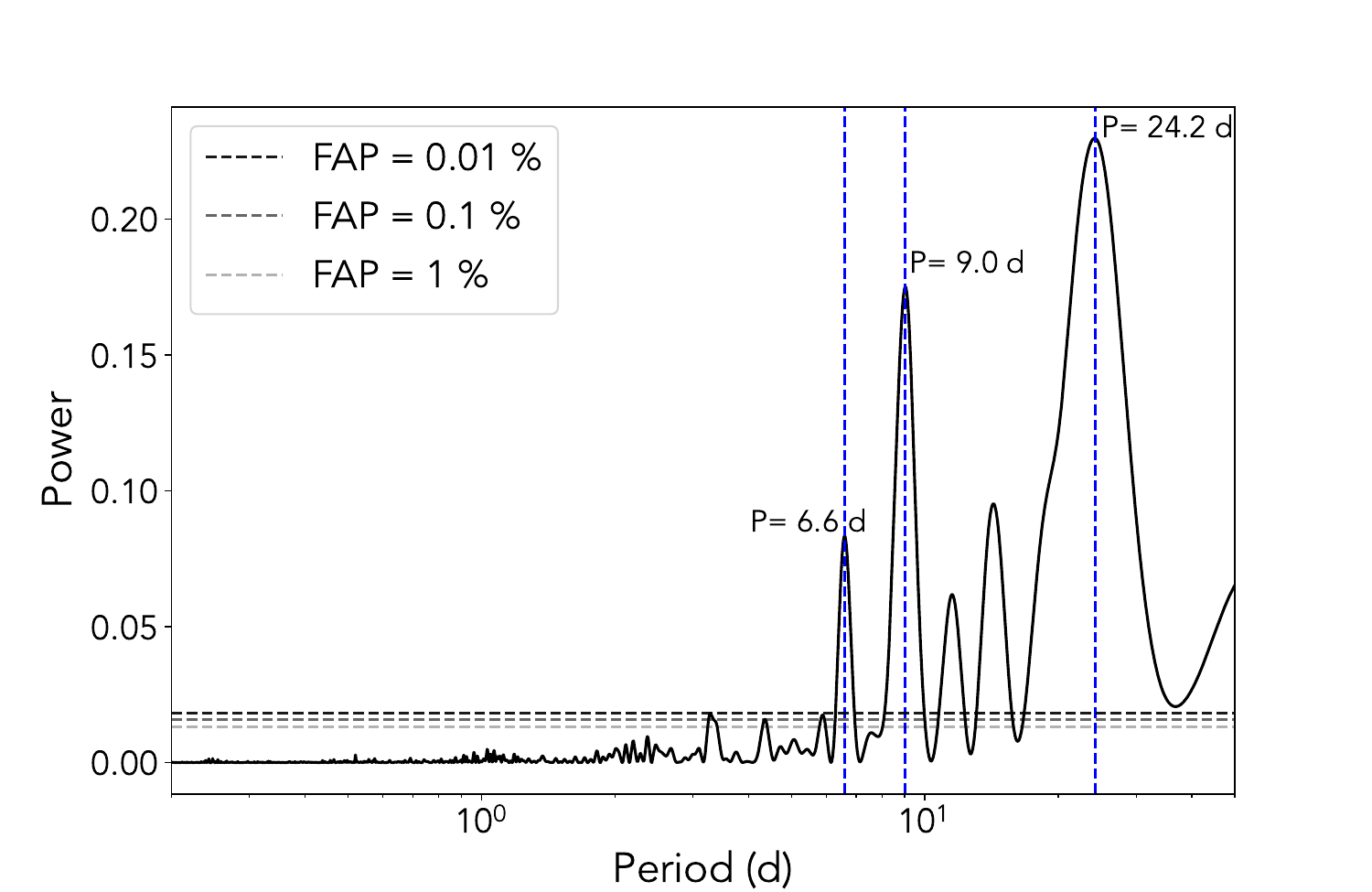} \\
   \includegraphics[width=9.5cm]{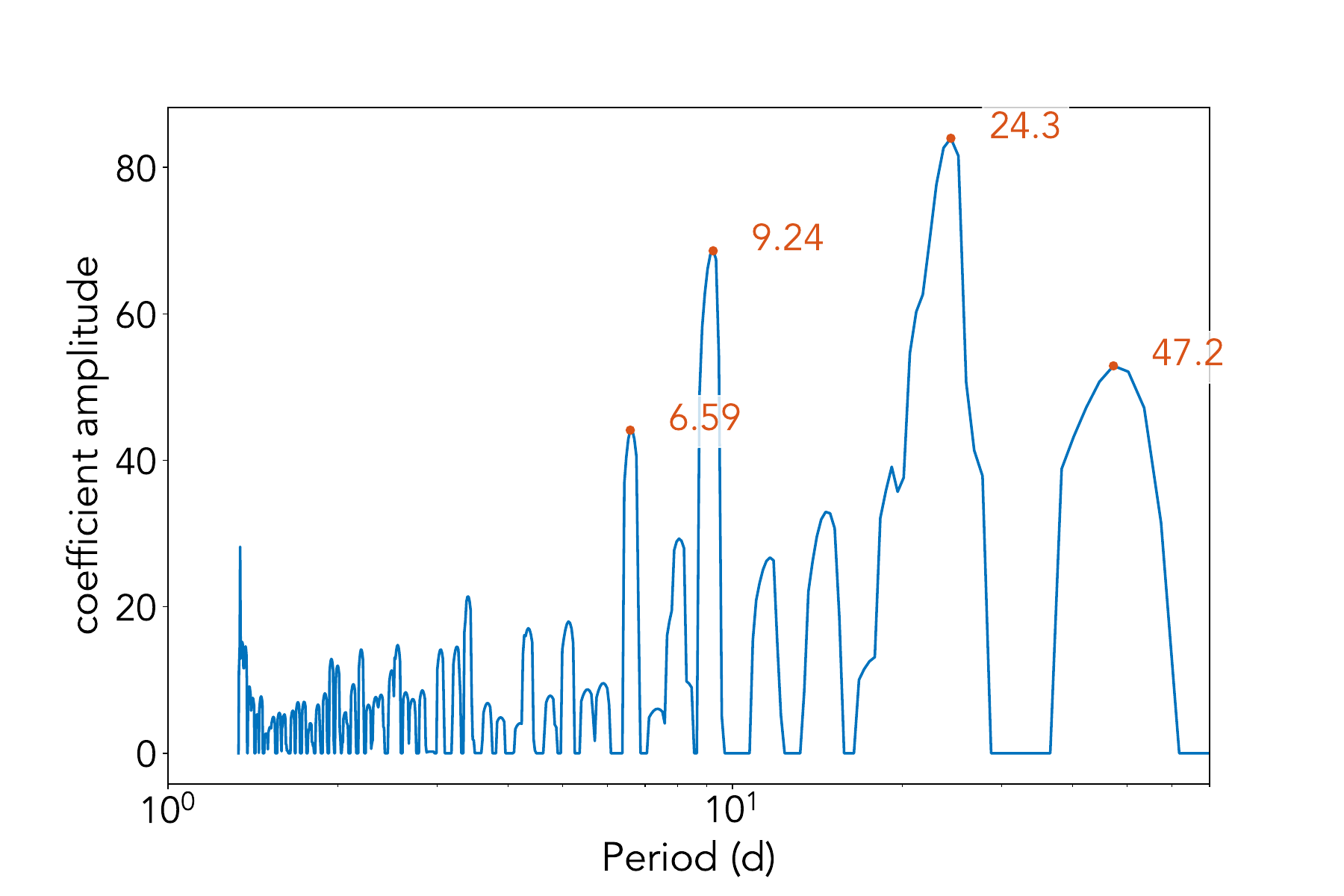}
      \caption{Top: K2 light curve of CI Tau. Middle: GLS periodogram of the K2 light curve, showing three peaks at 6.6\,d, 9.0\,d and 24.2\,d labeled with blue dashed lines. Bottom: L1 periodogram, showing the same periods as found in the GLS periodogram, with an additional significant peak at 47.2\,d}.
         \label{fig:k2lc}
   \end{figure}

The light curve of CI Tau (Figure \ref{fig:k2lc}: top panel) displays short-term erratic, aperiodic variations on time scales of hours to a few days. These are likely induced by surface inhomogeneities resulting from the combination of cool magnetic and hot accretion spots that appear and disappear on such time scales \citep{herbst1994, grankin2007, rucinski2008,findeisen2013}. Such variability is commonly identified as red noise in the data and will be explored further in Section \ref{section:rednoise_vaughan}. We computed a generalized Lomb-Scargle periodogram \citep[GLS,][]{Zechmeister2009}, as implemented in \textsc{Astropy} \citep{astropy2018}, to search for periods in the range $\sim$\,0.1 to\,100 d, using a frequency grid sufficiently dense to provide 50 samples across a given periodogram peak. The GLS periodogram is shown only for periods between 0.1 d and 50 d. We also include an L1 periodogram \citep{hara2017} of the K2 light curve in Figure \ref{fig:k2lc}, showing the most significant periods with their associated false alarm probabilities (FAP).

An L1 periodogram is a variant of the traditional Lomb-Scargle periodogram that uses the L1 norm as a measure of goodness-of-fit criterion. The key difference between the traditional Lomb-Scargle periodogram and the L1 periodogram is in the optimization criteria they use. While the traditional Lomb-Scargle periodogram uses the least squares method to minimize the sum of squared differences between the data and the fitted sinusoidal model, the L1 periodogram analyses multiple sinusoids simultaneously. Furthermore, the L1 periodogram proves particularly useful as a supplement to the Lomb-Scargle periodogram, offering a more lucid representation of the number of peaks and their significance. If for example the peaks identified by the L1-periodogram result in a chi-square value of the residuals consistent with periodogram noise, it suggests there may not be many additional signals and therefore helps avoid selecting spurious peaks.

The combined analysis using the GLS (and L1 periodogram), yields periodic variability with significant peaks at approximately 6.6\,$\pm$\,0.2\,d (6.59\,d), 9.0\,$\pm$\,0.5\,d (9.24\,d), 11.5\,$\pm$\,0.5\,d, 14.2\,$\pm$\,0.8\,d, and 24.2\,$\pm$\,5\,d (24.3\,d), as well as (47.2\,d). We compute $\log_{10}$ FAPS of -270, -196, -46 and -7 for the 24.3\,d, 6.59\,d, 9.24\,d and 47.2\,d periods found in the L1 periodogram, respectively.

\begin{figure}
  \centering
    \includegraphics[width=9cm]{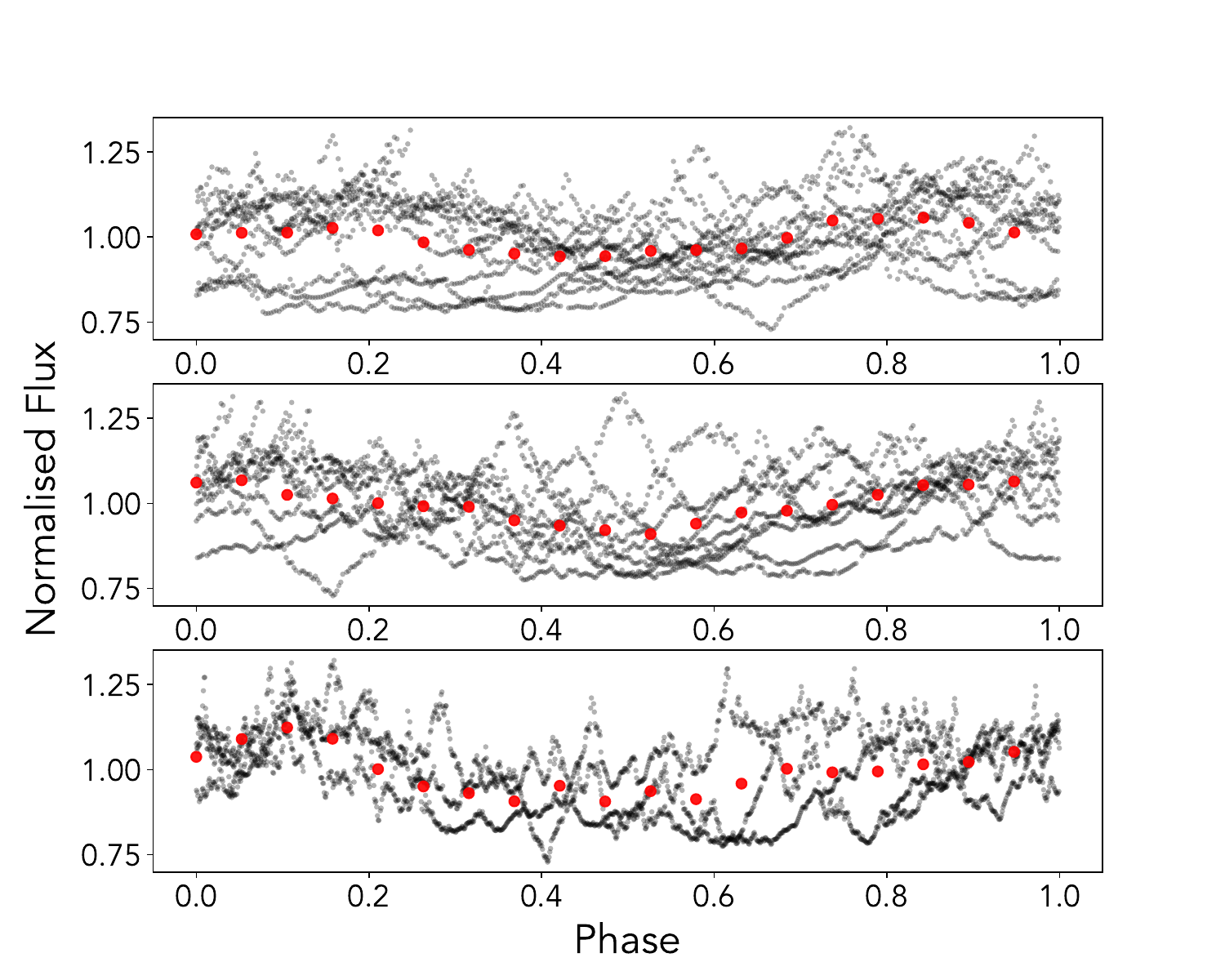}
      \caption{In black: K2 light curve phased on the 6.6\,d, 9.0\,d and 24.2\,d periods, respectively from top to bottom. The phase-binned mean light curve is shown in red.}
      \label{fig:k2_phased}
  \end{figure}

In Figure \ref{fig:k2_phased}, we present the K2 light curve phased on the dominant period peaks at 6.6\,d, 9.0\,d, and 24.2\,d. Errors on the periods listed for the GLS periodogram were obtained by considering the full-width at half-maximum (FWHM) of the individual periodogram peaks assuming they are Gaussians.

\citet{biddle2021} investigated the possibility of the 24.2\,d period being a synodic period of any combinations of the 6.6\,d, 9.0\,d and the 14.2\,d periods. The authors combined sinusoidal functions with periods of 6.6 and 9.0 days, and 14.2 and 9.0 days respectively, using the same sampling as the K2 data. None of the combinations were able to generate a peak comparable to the one observed in CI Tau with a period of $\sim$\,24.2\,d and therefore, it is unlikely that the 24.2\,d peak corresponds to a synodic period. We revisit the beat frequency (\(f_{\text{beat}}\)) calculations from combinations of any two significant frequency peaks ($f_1$ and $f_2$) that we find in the GLS periodogram, using: \( f_{\text{\rm beat}} = |f_2 - f_1| \). We find that the 14.2\,d period is likely a beat period of the 24.2\,d and the 9.0\,d periods. While the 11.5\,d period could be due to beating effects between the 6.6 d and 14.2\,d peaks.

The nature of the 6.6\,d and the 9.0\,d periods have been debated in the literature. \citet{johnskrull2016} and later \citet{biddle2021} claim that the 6.6\,d period corresponds to the stellar rotation period and the 9.0\,d is likely induced by a $\sim$\,11\,M$_{\rm Jup}$ planet. This view was later refuted by \citet[][Donati et al. 2023, submitted]{donati2020}, where ZDI analysis of ESPaDOnS spectropolarimetric data demonstrate clearly that the $\sim$ 9.0\,d period is due to spot rotational modulation. In this paper, we focus on the nature of the longer term variability, mainly the highest peak seen in the K2 light curve at 24.2\,d. Note that, henceforth in the text, we conveniently refer to this period identified in the K2 light curve as 24\,d.

\subsection{LCOGT: multi-wavelength optical photometry} \label{section:lcogt}

\begin{figure*}
  \centering
    \includegraphics[width=18cm]{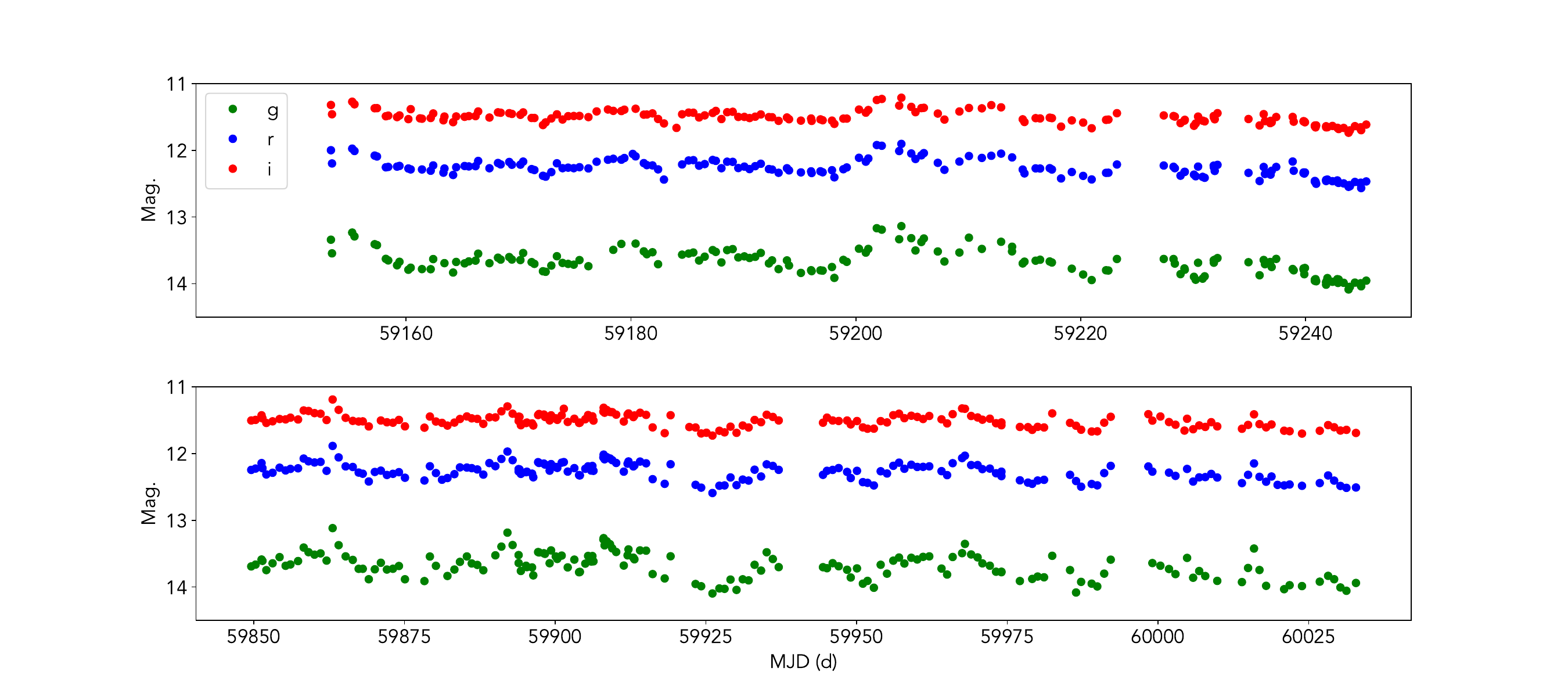} \\
    \includegraphics[width=18cm]{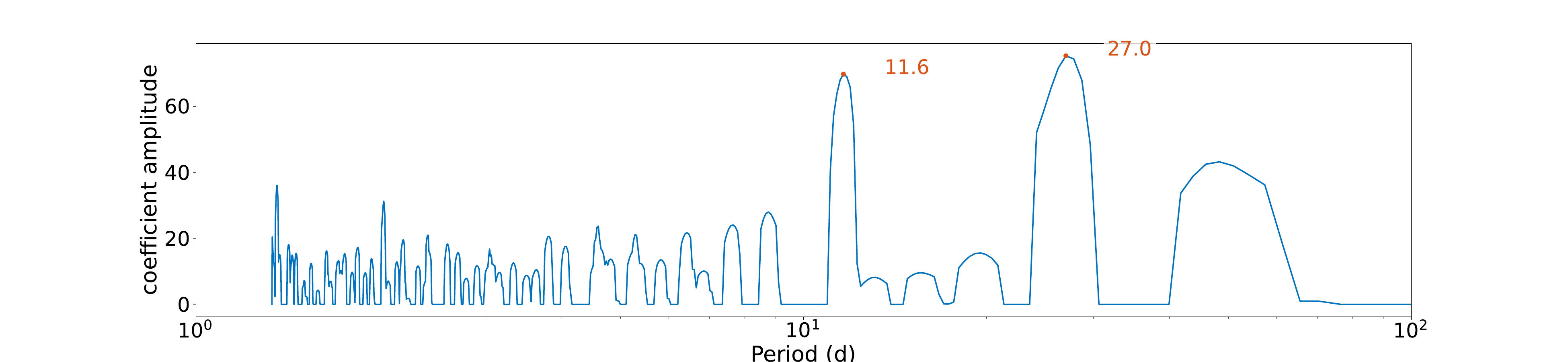} \\
        \includegraphics[width=18cm]{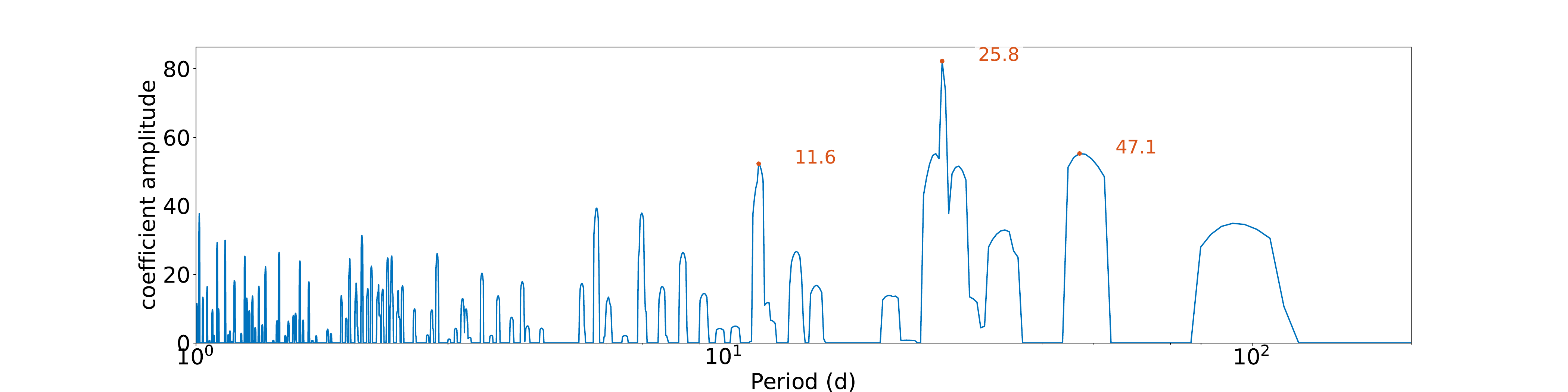} \\
      \caption{LCOGT light curves and L1 periogogram for each band, 'g' (green), 'r' (blue) and 'i' (red). Top panel: first semester (20B) light curve observed between 31 Oct. 2020 and 08 Jan 2021. Second from top panel: second semester light curve taken between 27 Sept. 2022 and 31 Mar. 2023. Third from top panel: L1 periodogram of combined data set for semester 20B. Bottom panel: L1 periodogram of combined data set for semester 22B/23A.}
      \label{figure:lcogt}
  \end{figure*}

The resulting g’r’i’ light curves of CI Tau obtained for the first and second epochs are shown in the top two panels of Figure \ref{figure:lcogt}\footnote{The table of photometric measurements is available electronically at CDS, Strasbourg.}. For the first (second) epoch, the peak-to-peak amplitude of variability reaches 0.94 (0.98), 0.67 (0.71), and 0.52 (0.54) mag in the g’r’i’ filters, respectively. 

We compute an L1 periodogram of light curves from both semesters (Figure \ref{figure:lcogt}). We see clear peaks at 11.6 d and 27 d in semester 20B, with $\log_{10}$ FAPS of -22 and -24, respectively. While in semester 22B/23A, peaks are seen at 11.6 d,  25.8 d and 47.1 d, with $\log_{10}$ FAPS of -6, -20 and -23, respectively.. These periods are consistent within their respective errors with the 11.5\,$\pm$\,0.5 d, 24\,$\pm$\,5 d and 47.2 d peaks found in the K2 light curve. We choose to conveniently cite the long term stable period at 25.8/27 d seen in the LCOGT light curve as 26.4\,$\pm$0.6\,d, which represents the mean period derived from the two semesters, and the error indicating half the difference between the two periods. We highlight that GLS periodograms were generated for the LCOGT light curves as well, producing results similar to those obtained from the L1 periodogram. We therefore opt not to include results from the GLS periodogram in the paper, as done for the K2 light curve.

Interestingly, the 9.0\,d, 6.6\,d and 14.2\,d peaks are not detected in the LCOGT light curve but are present in the K2 data, although the former is well sampled to detect this period. To examine the possibility of a sampling issue, we down-sampled the K2 light curve to mimic a mean sampling of 0.215\,d as in the LCOGT light curve. We still recover the same periods in the binned K2 light curve as in the original which implies that this discrepancy in periods found in the two datasets is not due to the time sampling. In addition, the LCOGT data does not sample burst time scales which can be of order of less than a mean sampling of 0.215\,d, while the K2 data does. 

We note that the K2 data were taken around 3 to 6 years before the LCOGT data. A non-detection of the aforementioned periods in the LCOGT data might indicate that these periods are either transient or potentially drowned by other activity-related phenomena operating over a more extended period of several years. Alternatively, the physical processes responsible for these periods might have a lifespan shorter than the time span covered by the LCOGT light curves (3-6 months). These processes may occur randomly in phase within the given time frame, resulting in their lack of detection in the LCOGT periodogram.

\subsection{Periodogram-peak significance-test in red noise} \label{section:rednoise_vaughan}

A challenging problem in observational astrophysics has been to detect periodic or almost-periodic signals in noisy time series. The task becomes even more challenging when we aim at detecting significant periodic signals hidden in stellar activity (e.g. spot modulation combined with red noise).

\begin{figure}
  \centering
    \includegraphics[width=9.5cm]{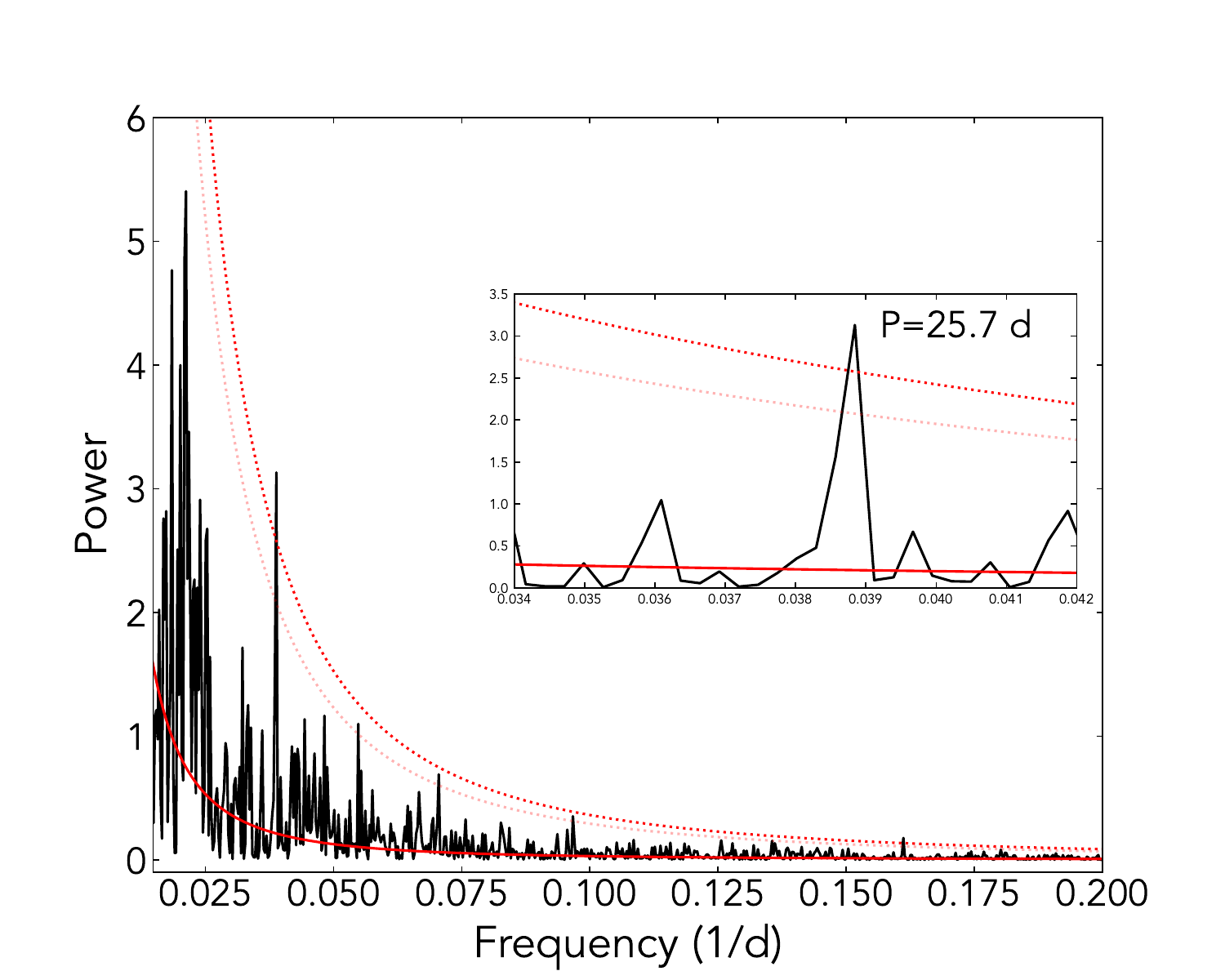}
      \caption{Power spectrum of the photometric time series (black line) fitted with a power law of the form $P(f) = Nf^{-\alpha}$ (red continuous line). The red dotted lines are at 0.1\% and 0.01\% FAP levels from bottom to top, respectively. The 25.7 d period is clearly significant above the 0.01\% FAP.}
         \label{fig:rednoi_pgm_vaughan}
  \end{figure}

As seen in Figure \ref{fig:rednoi_pgm_vaughan}, the power spectrum of CI Tau's K2 light curve (in black) consists of a red noise component which causes an increase in power at lower frequencies (red continuous line). To test the significance of the periodogram peaks in such red noise, we adopt the recipe of \citet{vaughan2005}. In particular we test the significance of the 24 d peak which is of main interest for this study. 

The method involves fitting the underlying power spectral density (PSD) shape of the classical periodogram \citep{schuster1898} and identifying any coherent, statistically significant peaks by computing false alarm probability (FAP) levels. The normalized periodogram is given by,

\begin{equation}
    I(f_j) = \frac{2 \Delta T}{{\langle x \rangle} ^2 N} |X_j| ^ 2,
\end{equation}

where $\Delta T$ is the sampling of the time step for an $n$ evenly sampled time series $x_n$. For the K2 light curve of CI Tau,  $\Delta T$ corresponds to  $\sim$ 30 min cadence. The periodogram \citep{schuster1898} of the time series is the modulus-squared of the discrete Fourier
transform, $X_j$, at each of the $\frac{n}{2} - 1$ independent Fourier frequencies, with $N$ being a normalisation factor. Only $\frac{n}{2} - 1$ frequency points are considered untill the Nyquist frequency, and frequencies higher than the Nyquist are ignored.

We note that since the classical periodogram of the K2 light curve (Figure \ref{fig:rednoi_pgm_vaughan}) is a variation of the GLS periodogram computed in Section \ref{section:GLS_periodogram}, we do expect slight differences between periodogram peaks obtained from the two methods, but in practice the classical periodogram is accurate enough to give useful insight into important periodic features \citep[e.g.][]{vanderplas2018}. For example, the 24 $\pm$ 5 d period peak obtained using the GLS method (Figure \ref{fig:k2lc}) is fully consistent within errors with the 25.7 d peak found in the classical periodogram (Figure \ref{fig:rednoi_pgm_vaughan}).

At any given frequency $f_j$, the discrete Fourier transform $I(f_j)$ is scattered around the true power spectrum, $P(f_j)$, following a $\chi^2$ distribution with two degrees of freedom, the mean and variance:

\begin{equation}
    I(f_j) = P(f_j) \chi^2 / 2,
    \label{eq: dft}
\end{equation}

where, $\chi^2$ has an exponential probability density (with a mean of 2 and variance of 4) of the form:

\begin{equation}
    {\rm PDF}_{\chi^2}(x) = \frac{1}{2}e^{-x/2}
    \label{eq:plpdf}
\end{equation}

We assume that the underlying power spectrum of the time series follows a power law of the form $P(f) = Nf^{-\alpha}$. We use a least-squares (LS) fitting method to fit the continuum of the log-periodogram in linear-space and extract $\hat{N}$ and $\hat{{\alpha}}$, which are parameters governing the red noise continuum model of the form $\hat{P}(f) = \hat{N}f^{-\hat{\alpha}}$ (see \citet{vaughan2005} for full description of the process). We use the null hypothesis that the data was generated by a process with the model spectrum and no periodic component. 

To infer the statistical significance of the 25.7 d peak, we use an analytic significance test based on the $\chi^2$ statistics, accounting for the number of frequencies sampled. Since the residual, $\gamma({f_j})$ = 2$I(f_j)/P(f_j)$ is $\chi^2$ distributed, the integrated area under the probability density function of the $\chi^2$ distribution (Equation \ref{eq:plpdf}) gives the probability that the power associated with a given peak is significant to a limit (1-$\epsilon$), where $\epsilon$ is the FAP level. When considered in units of the periodogram power, it is given by,

\begin{equation}
    \gamma_\epsilon = -2 \ln \big[ 1 - (1 - \epsilon) ^ \frac{1}{\frac{n}{2} - 1}\big],
    \label{eq:significance}
\end{equation}

Once $\epsilon$ is specified, $\gamma_\epsilon$ is calculated to give the significance level used to identify
outliers in the periodogram that indicate the presence of statistically significant peaks based on the FAP.

We compute the 0.1\% and 0.01\% FAP levels (red dotted lines, in Figure \ref{fig:rednoi_pgm_vaughan}). We find that the 25.7 d period peak in the classical periodogram (corresponding to the 24 $\pm$ 5 d in the GLS periodogram) is significant beyond the 0.01\% level (see inset panel of Figure \ref{fig:rednoi_pgm_vaughan}) and therefore is a significantly true period. 


\subsection{Gaussian process regression using a Quasi-periodic Kernel}

Rotationally modulated phenomena such as stellar spots, plages and faculae, induce signals in both radial velocities \citep[e.g.][]{rajpaul2015} and photometric time series \citep[e.g.][]{notsu2013}. If these phenomena are regular and stable over a long period of time with a given periodicity, we expect to be able to detect a clear periodicity related to these modulations in both RVs and photometry. However, stellar spots and faculae are not static, but rather evolve over time, emerging, changing shape, and have a finite lifetime. This results into non-periodic/quasi-periodic (correlated-noise-like) variations over each cycle of the stellar rotation, with shape and amplitude of the variability changing from cycle to cycle as seen in CI Tau's light curve. Inferring rotation periods using straightforward sinusoidal variability models are not suitable in such cases since these only account for white noise. An alternative is to use a model that employs stochastic variability such as a Gaussian process \citep[GP,][]{angus2018}.

In summary, a GP is a stochastic process that captures the covariance between observational data points and allows for the modelling of correlated (red) noise.  A GP is built using a covariance matrix with individual observation variances as diagonal elements and each off-diagonal element describes the covariance between any two observations. A kernel function is used to determine the values of the off-diagonal elements, which encapsulates the nature of the correlated noise. GPs provide a great deal of flexibility that has made them an effective tool to account for stellar activity \citep{haywood2014}. We employ a GP regression method as prescribed by \citet{rajpaul2015} to model the effects of stellar activity in the data (spots and random processes) that enables us to gauge the robustness of the 24.2 d period in the light curve. The recommended kernel for stellar activity characterization is 
a quasi-periodic (QP) kernel \citep{aigrain2012} of the form:

\begin{equation}
\label{eq:cov_qp}
 {\gamma _{qp}(t,t')} = h^ 2 \exp \left\{ - \frac{\sin^2 \left[\pi (t-t')/P\right]}{ 2 \lambda_{\rm p}^2} 
 - \frac{(t-t')^2}{2\lambda_{\rm e}^2} \right\},
\end{equation}

The QP kernel uses four parameters (termed as hyperparameters) to model the stellar activity: $h$, $P$, $\lambda_{\rm p}$ and $\lambda_{\rm e}$. $t$ and $t'$ are times of observations made at subsequent intervals. 
$P$ and $\lambda_{\rm p}$ represent the period and smoothing parameter of the periodic component of the variations. The parameter $\lambda_{\rm e}$ denotes the evolutionary time scale. The parameter $h$ regulates the GP model amplitude. Additionally, we incorporate a white noise term as $\sigma_{K2}^2 \delta(t, t')$, where $\sigma_{K2}$ represents the noise amplitude, and $\delta(t, t')$ is the Kronecker delta function.

We employ MCMC sampling to model both the 6.6\,d and 9.0\,d periods simultaneously, utilising a combination of two QP kernels, each corresponding to these distinct periods. The fitting process is performed using MCMC sampling, implemented through the Python package \textsc{emcee}. To explore parameter space, we employed a total of 256 walkers and executed 10,000 iterations during the MCMC sampling process. The initial positions for these walkers were set to correspond to initial parameter guesses. We guess $\lambda_{\rm e}$ through a visual examination of the data to identify a characteristic time scale governing variations in the time series - such as changes in shape, amplitude, and phase. Our estimation points towards a value of approximately 20 days which was fixed in the fitting process. For the remaining hyper-parameters, we assume uniform priors, as detailed in Table \ref{tab:tab_k2_gp}. The best-fit hyper-parameters and their corresponding errors are summarised in the same table.

\begin{table}[htbp]
     \caption{Priors and best fit parameters for MCMC fit of the QP-GP for the 6.6\,d and 9\,d periods.}
    \centering
   \renewcommand{\arraystretch}{1.3} 
    \begin{tabular}{l l l}
        \hline
        Parameter & Prior & Fit Value \\
        \hline
          \rule{0pt}{10pt} 
        \rule{0pt}{2pt}
        Hyper-parameters for 9\,d spot & & \\
         \rule{0pt}{10pt} 
        GP amplitude ($h_1$) [K2 mag.] & $\mathcal{U}$[0.05, 1.0] & $0.46_{-0.02}^{+0.8}$ \\
        Spot period ($P_1$) [d] & $\mathcal{U}$[8.5, 9.5] & $9.3_{-0.3}^{+0.2}$ \\
        Smoothing ($\lambda_{p{_1}}$) & $\mathcal{U}$[0.1, 1.5] & $0.2_{-0.07}^{+0.01}$\\
        Evol. time scale ($\lambda_{e{_1}}$) [d] & fixed & 20.0 \\
        White noise amplitude ($\sigma_{k2}$) & fixed & 0.01 \\
        \hdashline 
        Hyper-parameters for 6.6\,d period & & \\
        GP amplitude ($h_2$) [K2 mag.] & $\mathcal{U}$[0.05, 1.0] & $0.45_{-0.01}^{+0.7}$ \\
        Spot period ($P_2$) [d] & $\mathcal{U}$[6.0, 7.0] & $6.6_{-0.2}^{+0.2}$ \\
        Smoothing ($\lambda_{p{_2}}$) & $\mathcal{U}$[0.1, 1.5] & $0.8_{-0.03}^{+0.4}$\\
        Evol. time scale ($\lambda_{e{_2}}$) [d] & fixed & 20.0 \\
               \hdashline 
        Fit report & & \\
               RMS data [K2 mag.] & - & 1.0 \\ 
               RMS residuals [K2 mag.] & - & 0.09 \\
        \hline
    \end{tabular}
    \label{tab:tab_k2_gp}
\end{table}

\begin{figure}
  \centering
    \includegraphics[width=9.5cm]{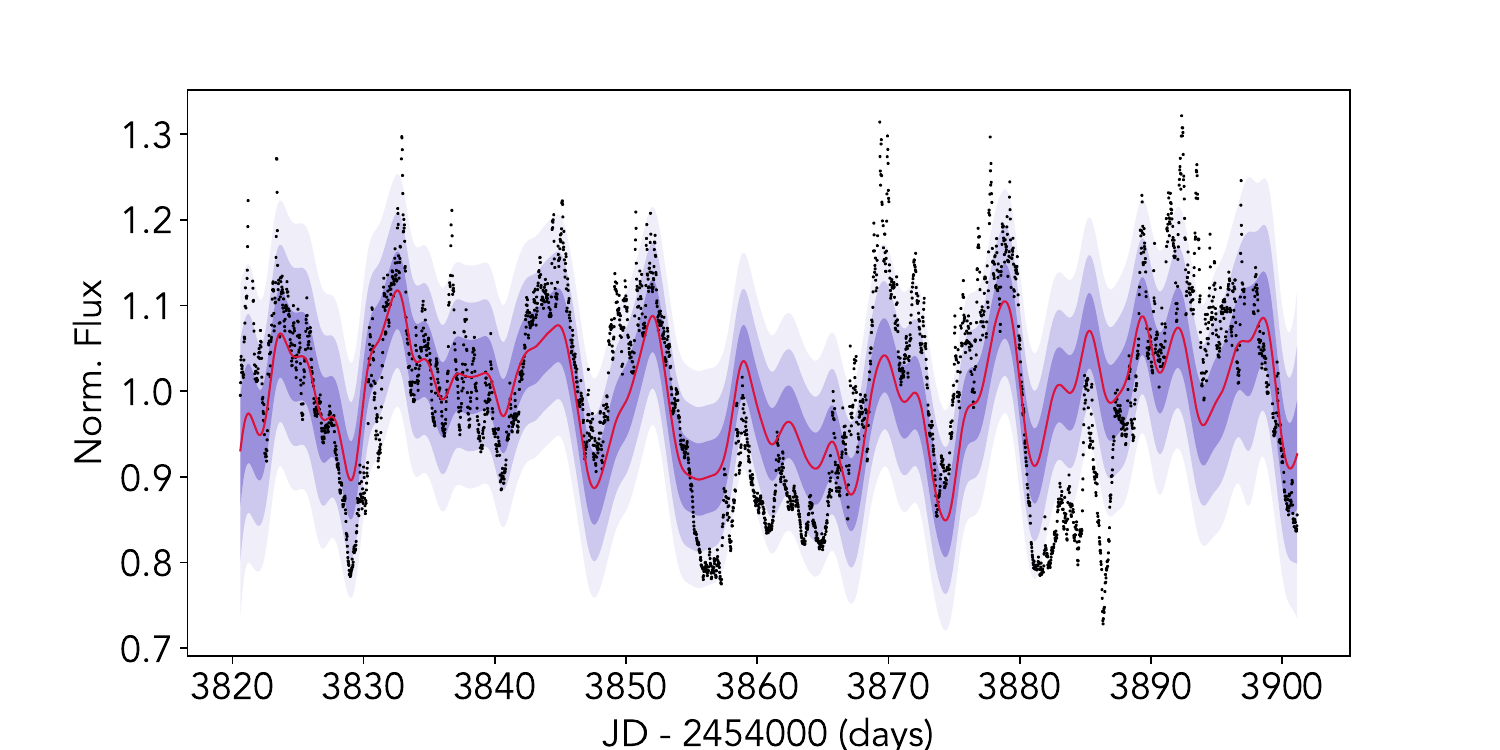} \\
     \includegraphics[width=9.5cm]{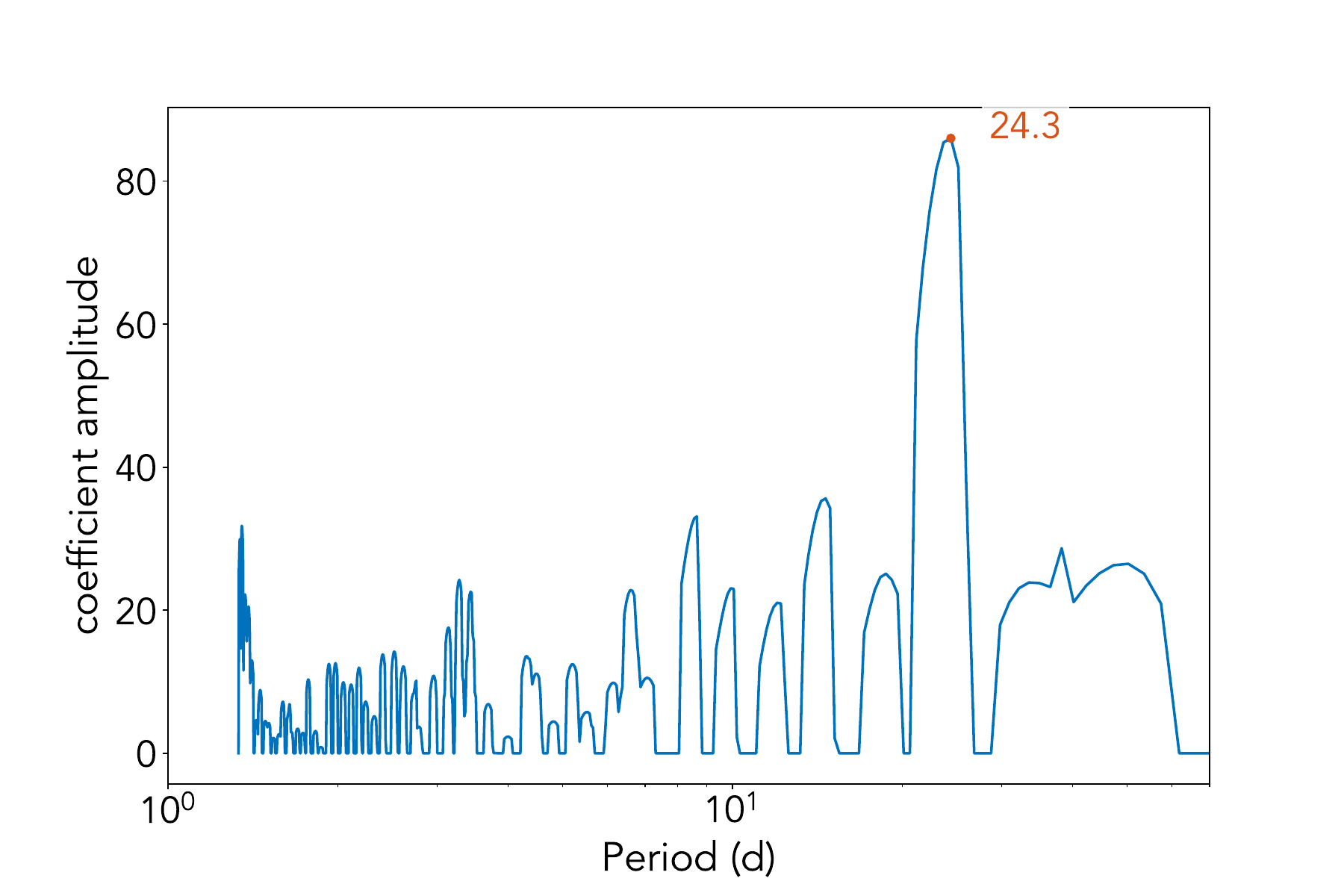} 
      \caption{A quasi-periodic GP model with 2 spots of periods 6.6 d and 9.0 d shown by the red dashed curve. The shaded region shows the 1-$\sigma$, 2-$\sigma$ and 3-$\sigma$ from the mean model. The bottom panel shows the L1 periodogram} of the residual light curve after removing the modulation due to stellar spots.
         \label{fig:gp_kepler}
  \end{figure}

The GP model is shown as a red curve in the top panel of Figure \ref{fig:gp_kepler}. The L1 periodogram of the residual light curve after removing the modelled periods still shows a clear peak at 24.3 d with a log$_{10} $ FAP of -58 (Figure \ref{fig:gp_kepler}, bottom panel). This indicates that the 24 d period is not related to any of the 2 periods at 6.6 d and 9.0 d periods, i.e. not a synodic period, consistent with the findings of \citet{biddle2021}. We also test the possibility of the 24 d period being the beat period between either of the 6.6/14.2 d and the 9.0 d. We simulate 1000 light curves for each period combination, with period peak amplitudes and red noise levels that mimic the original K2 light curve. These light curves are then summed by continuously phase-shifting one of them to cover all phases. We then compute FAP levels on the normalised powers using all simulations. The results suggest that at any given phase, adding any of these 2 periodic models does not give rise to a beat period peak in the periodogram at 24\,d as high as the one seen in the original K2 light curve.

\section{Spectroscopic Analysis} \label{section:rv_analysis}
\subsection{ESPaDOnS RV analysis} \label{sect:espadons_rvs}

\begin{figure}
  \centering
    \includegraphics[width=9.0cm]{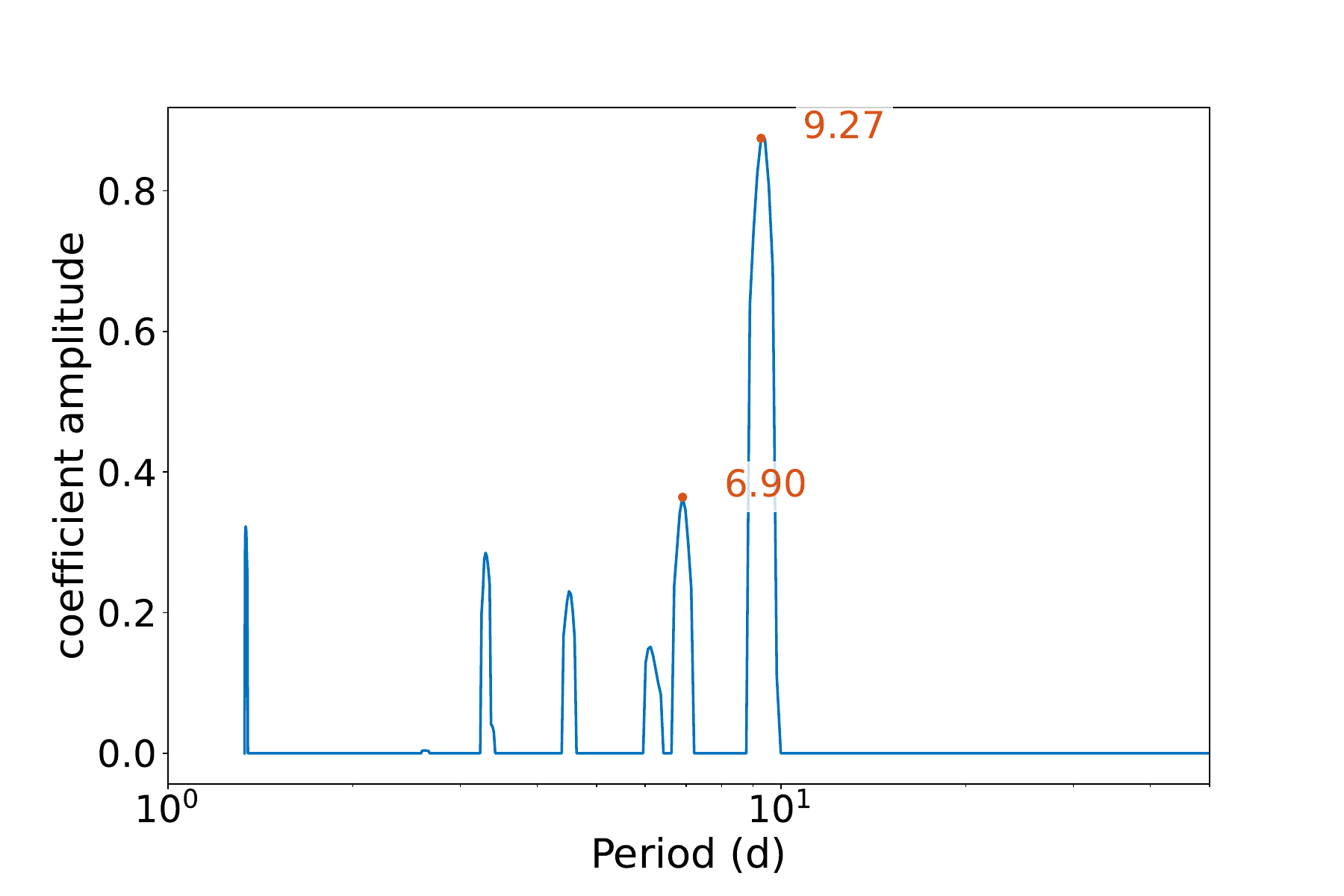}
      \caption{L1 periodogram of RVs related to 72 data points (4 nightly sub-exposures) from ESPaDOnS. The spot period is recovered at 9.27 d and another period is recovered at 6.9 d.}
         \label{figure:RVs_periodo_ESPADONS_full}
  \end{figure}

The RV measurements\footnote{The ESPaDOnS RVs will be available as a CDS table.} were obtained by fitting Gaussian functions to each mean LSD profile, using the python LMFIT package \citep{newville2016}.  An L1 periodogram of the full 72 RV data points is shown in Figure \ref{figure:RVs_periodo_ESPADONS_full}. Two significant periods in the whole ESPaDOnS RV dataset are seen at 9.27 d ($\log_{10}$ FAP = -11) and 6.9 d ($\log_{10}$ FAP = -10), both of which are also seen in the K2 light curve.

\begin{figure}
  \centering
    \includegraphics[width=9.5cm]{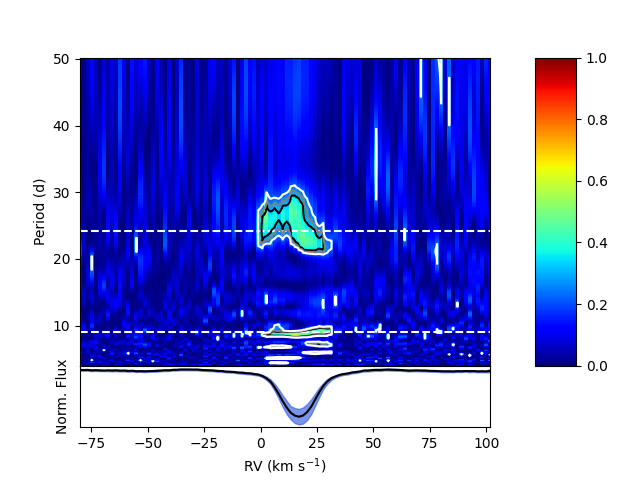}
      \caption{2d periodogram and mean LSD line profiles using 72 ESPaDOnS spectra. Top panel: 2d periodogram with colour range corresponding to the power of the
periodogram. The horizontal lines are drawn at 9.0 d and 24.2\,d. Contours represent FAP levels at 0.01\% (black), 0.1\% (grey) and 1\% (white). Bottom panel: Mean LSD profile in black, with 1 standard deviation of all profiles shaded in blue.}
         \label{fig:2d_espadons_periodo}
  \end{figure}

In addition, we show a 2D periodogram of the LSD profiles, for the 72 ESPaDOnS spectra in Figure \ref{fig:2d_espadons_periodo}. The 2D periodogram is useful to discern periodicities in each velocity channel of the LSD profile. To achieve this, we divided the line profiles into smaller velocity bins. Subsequently, periodograms were independently computed for each velocity bin. The resulting individual periodograms are visualized in a 2D representation as a function of each velocity bin, with the power of the periodogram depicted by colours \citep[e.g.][]{alencar2002,sousa2021}. From Figure \ref{fig:2d_espadons_periodo} it is clear that both the 9 d and the 24 d periods seen in the K2 light curve, are also seen to be present in the LSD mean profiles of ESPaDOnS.

\subsection{ESPaDOnS Bisector analysis} \label{section:bisector}
Stellar activity such as spots and faculae induce variability in photospheric lines, which in turn has an effect on RV measurements \citep[e.g.][]{reiners2013}. A first-order effect is seen when the spots occult either the receding or approaching half of the visible stellar hemisphere, modifying the shape of the spectral lines. This causes a apparent shift in the disk-averaged RV \citep[see e.g.][]{donati2020} that can mimic exoplanet reflex-motion signatures, adding confusions to a true Doppler signal due to a planetary companion that may be present in the system \citep[][]{queloz2001,desidera2004, huelamo2008, carolo2014}.

The bisector span (BIS span) measures the asymmetry of the mean line profile as was shown in a seminal study by \citet{queloz2001}. This method is a well known proxy to help gauge whether observed RV variations are due to stellar activity \citep[e.g.][]{queloz2001,meunier2010, dumusque2014, donati2020} or otherwise likely induced by Doppler reflex-motion signal by, e.g a planet \citep{moutou2005,naef2007}. If the asymmetry of the mean line profile is purely anti-correlated with the estimated RV, then it is a sign that the line profiles are being modulated by spot activity only. Otherwise, a flat BIS span with respect to the estimated RV indicates that the shift in the lines are purely due to Doppler reflex motion \citep{queloz2001}. In a mixed scenario when we have both a Doppler reflex by a planet and the effect of activity due to spot, both inducing RV signals that are comparable, we expect both an anti-correlation and a translational horizontal scatter of the BIS span \citep[see][]{simola2019} and see also Section \ref{sec:spirou_bis}, for a detailed analysis).

\begin{figure}
  \centering
    \includegraphics[width=9.5cm]{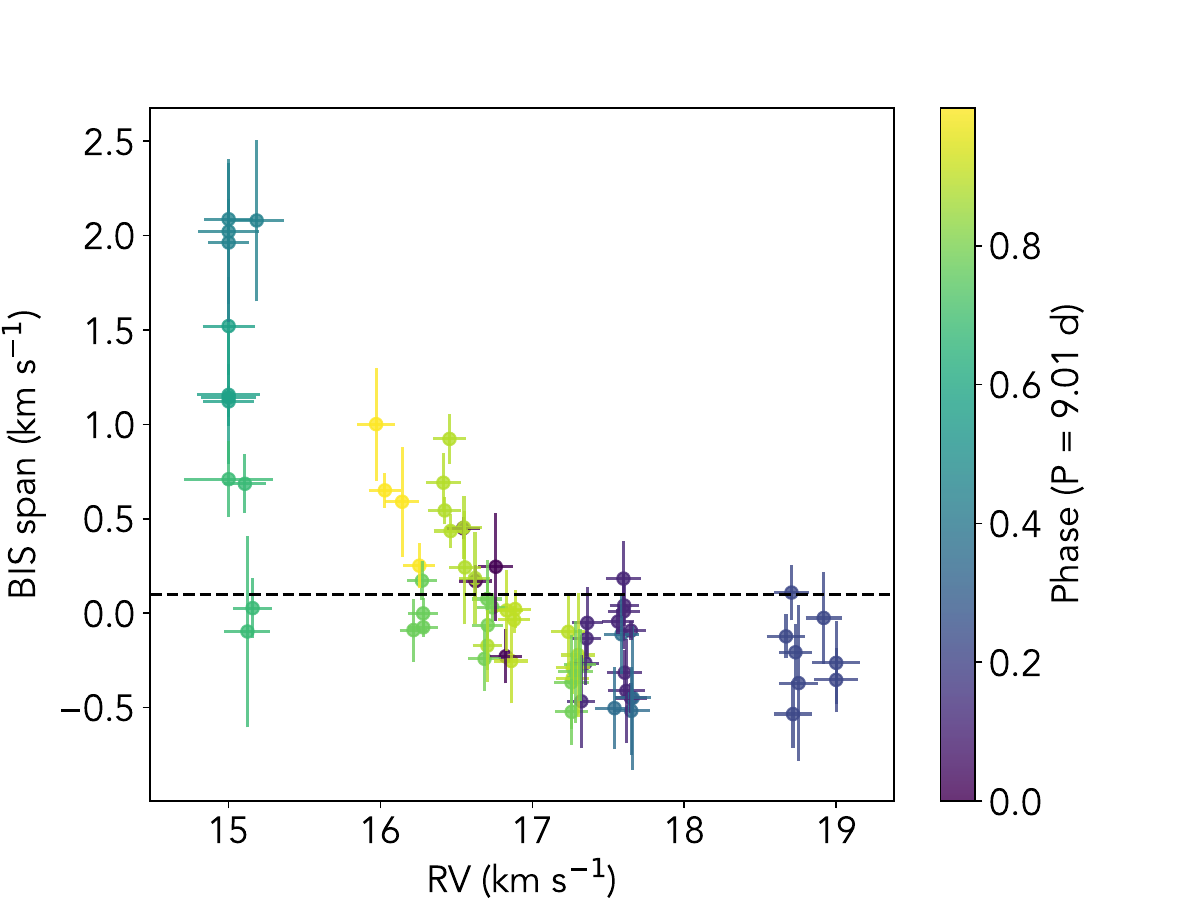}
      \caption{Bisector span of the mean LSD profiles from ESPaDOnS spectra. A correlation is seen at lower velocities up to around 17 km s$^{-1}$ and then no correlation where it reached a plateau. The data points are colour-coded based in phase of the 9.01 d spot rotation period. The dashed line indicates the threshold below which BIS span values were selected to compute the L1 periodogram shown in red in Figure \ref{fig:bis_periodo}.}
         \label{fig:bis}
  \end{figure}

We carried out a BIS span analysis using the mean LSD profiles of CI Tau. The ESPaDOnS data is of particular interest because at optical wavelengths, the contrast between the spot and the stellar continuum is largest, which makes any stellar activity due to spots more detectable. We followed the standard recipe for BIS span computation in \citet{queloz2001}. We divided each mean LSD profile in 100 bisector vertical intervals and computed the mean of each interval to obtain the bisector. The mean velocity of the top ($\overline{v_t}$) and bottom ($\overline{v_b}$) 10\%  was computed to obtain the BIS span given by $\overline{v_t}$ - $\overline{v_b}$ \citep[see][for details]{queloz2001}. A plot of the BIS span as a function of the RVs derived from the LSD profiles is shown in Figure \ref{fig:bis}.

We find that there is an anti-correlation between the BIS span and RVs at velocities lower than $\lesssim 17.0$ km s$^{-1}$, after which the BIS span reaches a plateau (Figure \ref{fig:bis}). The data points are colour-coded according to the phase using the ephemeris: BJD0 (d) = 2457736.76 + 9.01$c$, where $c$ denotes the spot rotation cycle, starting from an arbitrary initial BJD0 of 2457736.76, which corresponds to the first ESPaDOnS observation.

The BIS span is observed to be dependent on the phase of the spot rotation period at 9.0 days, consistent with findings reported by \citet{donati2020} and not very different from the newest value of 9.01\,d reported in Donati et al. 2023 (submitted). Between phases $\sim$ 0.5 to around 1.0, the spot seems to be in a configuration in which it induces a noticeable effect on the overall mean line-shape, causing an apparent RV shift which is intrinsic to the star's spot activity. During the rest of the phases, the effect of spot activity becomes minimal and the BIS span is flat. A flat BIS span suggests no correlation with RVs and therefore RV shifts of non-intrinsic nature \citep[e.g.][]{udry2000}. CI Tau's BIS span therefore indicates a mixed scenario whereby stellar activity due to spot, is seen, but there is also a second component not related to spot modulation, where pure RV shifts are detected in the lines.

    \begin{figure}
  \centering
    \includegraphics[width=9cm]{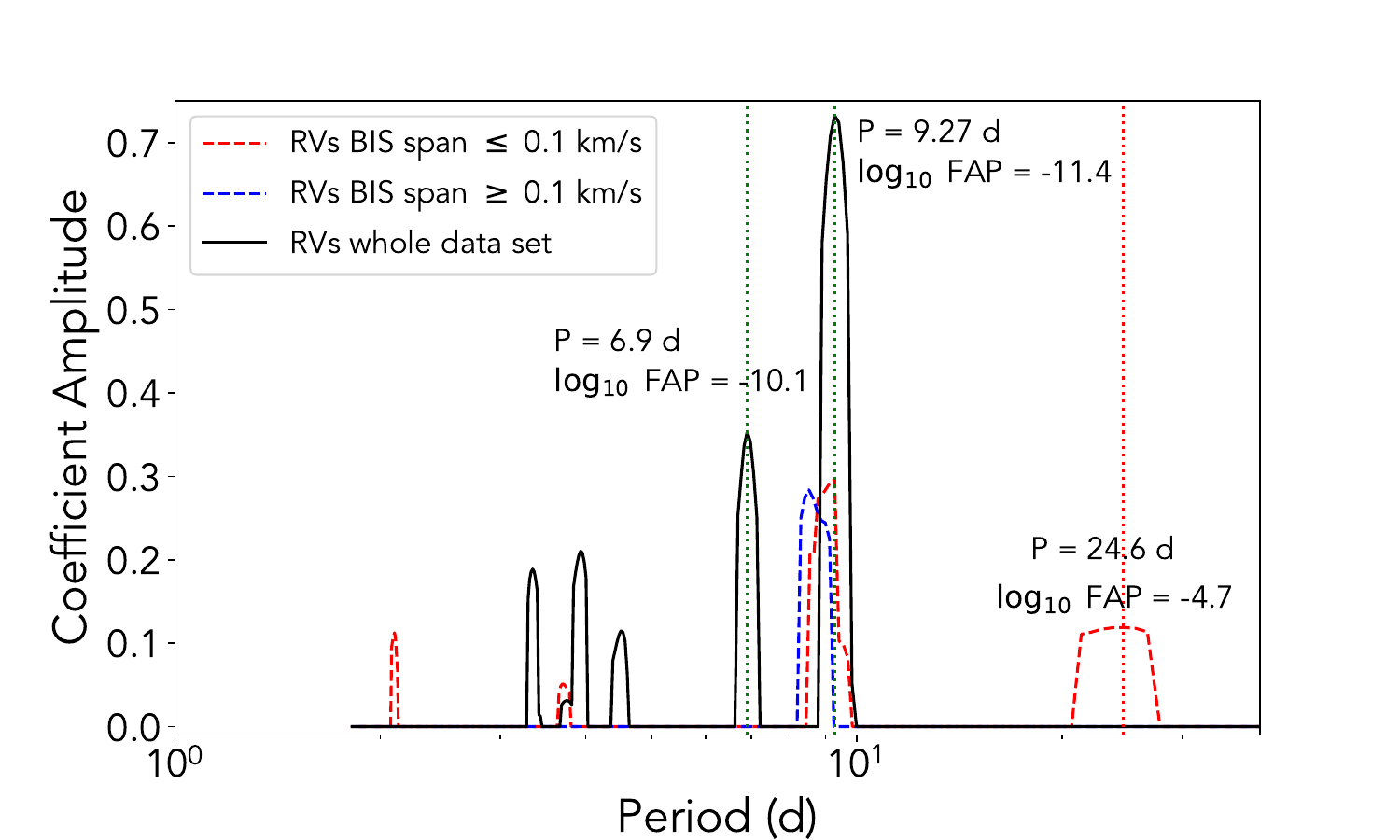}
      \caption{L1 Periodogram of RVs for the whole ESPaDOnS RV dataset (black), RVs selected based on BIS span values of $\geq$ 0.1 km s$^{-1}$ (blue) and RVs selected based on BIS span values of $\leq$ 0.1 km s$^{-1}$ (red). The level of activity due to stellar spot, corresponding to 6.9 d and 9.27 d (vertical green-dotted lines) is high in the whole RV dataset and RVs corresponding to BIS Span values of $\gtrsim$ 0.1. For the flatter part of the BIS span with values $\lesssim$ 0.1, the RVs display a clear period peak at $\sim$ 24.6 d (vertical red dotted line).}
         \label{fig:bis_periodo}
  \end{figure}

In Figure \ref{fig:bis_periodo}, we display the L1 periodogram of the whole RV dataset and RVs related to BIS span values below and above 0.1 km s$^{-1}$, which contain 45 and 27 data points, respectively. We arbitrarily chose this threshold because below this limit the BIS span is flat, although we could not exclude some contamination from spot modulation at RVs in the range $\sim$ 16.2 to 17.0 km s$^{-1}$. Our selection criteria results in 45 RV data points span over 64 days, which correspond to BIS span values that are essentially in the flatter part of the BIS span plot. We notice that the $\sim$ 24.6 d period peak is suppressed during spot activity, but rises to a significant FAP level, when RVs corresponding to only the flat part of the BIS span are considered.


\subsection{SPIRou RV analysis}

Activity signatures induced by spots are known to be strongly chromatic
\citep[e.g.][]{reiners2010}, since the spot-to-stellar continuum contrast tends to be lower at longer wavelengths in the NIR. In a recent study by \citet{miyakawa2021}, it was shown that chromaticity can actually be even more pronounced if we consider wavelengths from the visible ($K_p$, Kepler band covering 400 nm to 900 nm) to the NIR. \citet{miyakawa2021} studied RV jitter due to stellar activity on RVs in four different passbands ($K_p$, $J$, $H$ and $K_s$). The authors found that RV 'jitter' can vary from  1.4 km s$^{-1}$ down to 0.16 km s$^{-1}$ from $K_p$ to $K_s$ bands. This has important implications, especially for planet searches via spectroscopy in very young and active stars like CI Tau, where the effects due to spots may induce apparent RV shifts in the CCFs that are comparable to or even higher than pure Doppler RV shifts that can be induced by a planetary companion (see Section \ref{sec:spirou_bis}).

We utilised the \textsc{spirou-ccf} package, which employs a cross-correlation method to determine RVs\footnote{The SPIRou RVs will be available as a CDS table.} . The input data for our analysis are the telluric-corrected spectrum, which are calculated using the APERO pipeline \citep{cook2022}. The \textsc{spirou-ccf} package includes several processing steps to reduce the significant systematic effects that are commonly found in near-infrared (NIR) data. The cross-correlation process has been described in detail in \citet{martioli2022}. 

To test the chromatic effect of the spot on the RVs, we explored 2 separate cases. In the first case, we constructed a template spectrum (T1), adapted to the spectral type of CI Tau. To generate a template, we formulate a mask by selecting spectral features based on their proximity to the spectral type of the star. Each mask comprises a collection of around 1844 atomic and molecular lines in the whole SPIRou domain, with central wavelengths sourced from the VALD catalog \citep{piskunov1995}. The line depths are empirically derived from the template spectra of luminous stars observed by SPIRou.

\begin{figure}
  \centering
    \includegraphics[width=9.5cm]{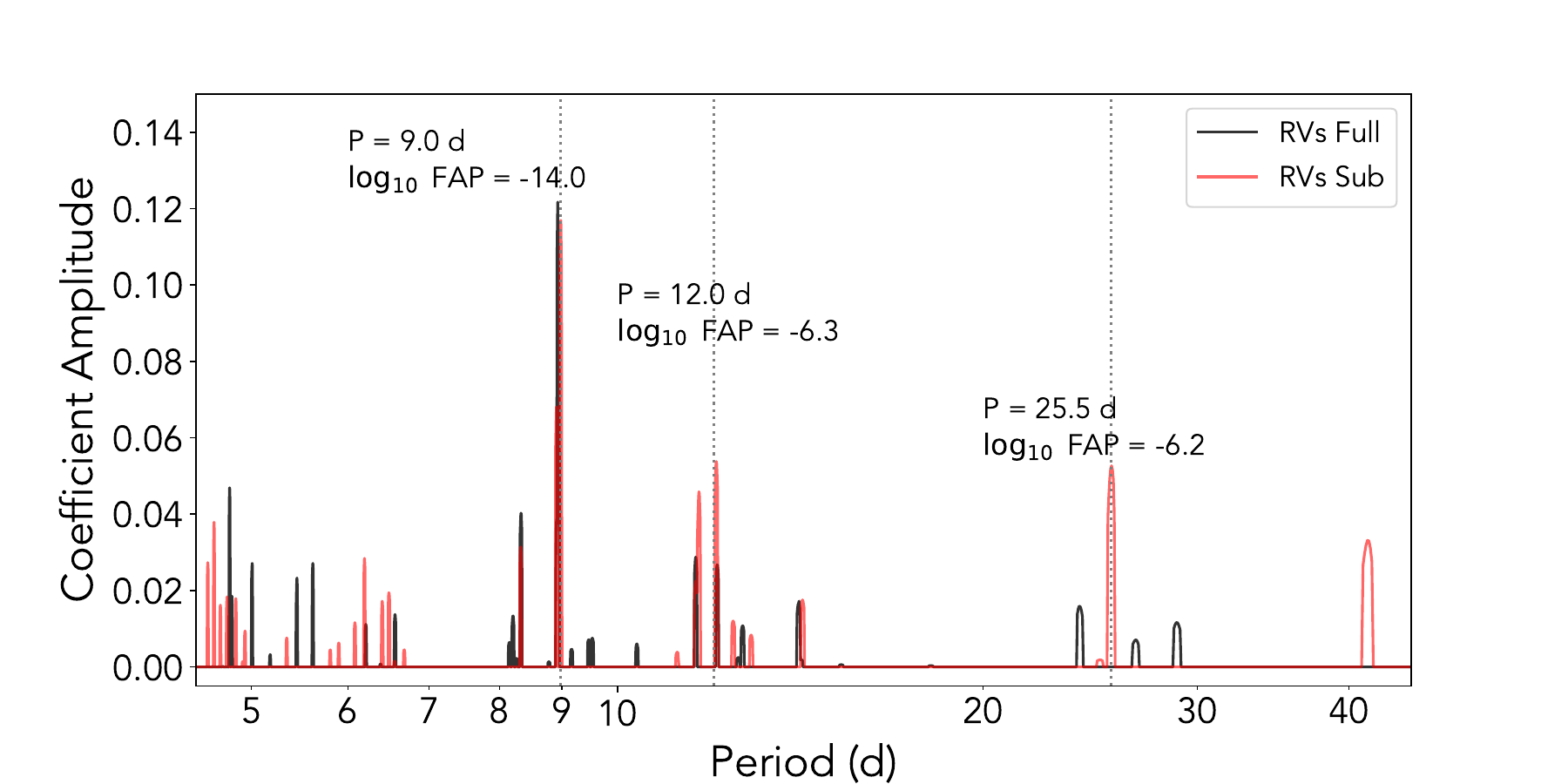}
      \caption{L1 periodogram of CCF RVs extracted from the full SPIRou wavelength range (in black) and CCF RVs related to the 2293 - 2340 nm wavelength range (in red). The dotted vertical lines mark the significant peaks based on their FAP, one of them being the 25.5 d peak.}
         \label{fig:gls_rvs_compa}
  \end{figure}

Each individual telluric-corrected spectra were then cross-correlated with T1 to obtain mean cross-correlation function (CCF) profiles of the absorption lines from which the RVs (which we refer to as RVs$_{full}$) were extracted. For the second case, to make sure we are least affected by spot activity, we considered a wavelength range as red as possible in the SPIRou domain, making sure we had enough photospheric lines for CCF computation and also making sure that there are not magnetically sensitive lines \citep[e.g.][]{lavail2019}. Upon close inspection, we found the 2293 - 2340\,nm range in the CO band matched this selection criteria the best. We constructed a sub-template (T2) from T1, over the 2293 - 2340\,nm wavelength range. We then cross-correlated all spectra with T2 and the RVs (which we refer to as RVs$_{sub}$) were extracted from this domain separately. Figure \ref{fig:gls_rvs_compa} shows the L1 periodogram of the CCF RVs for RVs$_{full}$ and RVs$_{sub}$. It can be seen that the peak corresponding to the 25.2 d period is significant for RVs$_{sub}$ compared to RVs$_{full}$. \footnote{On the longer term, we find two other peaks at 91.2 d and 413 d in the SPIRou CO RVs.}

Our analysis aligns well with the results of \citet{reiners2010, miyakawa2021}, which outline the chromaticity of the spot signal in the RVs. Therefore, to make sure we are least affected by this effect, we carried out our SPIRou orbital analysis considering RVs extracted over the 2293 - 2340\,nm range. To make sure that the chosen lines in this region are photospheric, we determined the veiling value of CI Tau for a given spectrum ($S$) as described in \citet{sousa2023}. We then computed the residual spectrum, by subtracting $S$ from a template veiled to the veiling value we computed in the K-band. The residual profiles show no feature at the location of the photospheric lines, which indicates that the lines are purely photospheric. A plot of all the CCF CO line profiles and their mean is shown in Figure \ref{fig:spirou_ccf_profiles}.


\subsection{SPIRou bisector analysis} \label{sec:spirou_bis}
We use the CCFs obtained by cross-correlating the full SPIRou wavelength range to compute the bisector span (BIS span) values as we described in Section \ref{section:bisector}. A Spearman correlation test on the SPIRou RVs against the obtained BIS values, yields an intermediate anti-correlation coefficient of -0.47 $\pm$ 0.05. The correlation coefficient's uncertainty is obtained using a Monte-Carlo simulation on the individual RVs and BIS values including their errors. 


In an attempt to investigate what scenario would give rise to the observed SPIRou bisector span, we have devised a simple model to demonstrate the impact of a spot on the bisector under two circumstances: case 1) when the central star remains stationary and case 2) when it undergoes a Doppler shift caused by the presence of a companion. In our model, we conduct experiments to investigate how the presence of a spot affects the disk-averaged star CCF and its impact on the bisector. We make the assumption that both the star and spot CCF are Gaussians, denoted by G$_{\rm star}$ and G$_{\rm spot}$, respectively. We take a realistic approach by considering G$_{\rm star}$ to be the mean profile of all available SPIRou CCFs. To achieve that, we fit Gaussians to all of the available 425 SPIRou mean CCF profiles. The Gaussian fit parameters, amplitude ($A_{star}$), mean ($\mu_{star}$) and standard deviation ($\sigma_{star}$), together with the associated error in the respective fit parameters are recorded. 

As for G$_{\rm spot}$, it has been studied in detail by \citet{donati2020} through Zeeman Doppler imaging (ZDI). It was illustrated that CI Tau's surface contains an accretion region characterized by a bright chromospheric region that aligns spatially with a dark and cool spot. \citet{donati2020} were able to constrain the spot's surface area to be approximately 20\% of the stellar surface. Since the impact of this cold spot on the disk-averaged spectrum is less pronounced than that of the unspotted portion of the stellar surface, the presence of a cool spot manifests as a distinctive feature, resembling a bump, which traverses the spectral line profile from the blue to the red end \citep{rosen2012,hebrard2014}. To simulate this bump, we incorporate the effect of the cold spot on the disk-averaged CCF of the star. We assume specific parameters for \textbf{G$_{\rm spot}$} that are, amplitude ($A_{spot}$), mean ($\mu_{spot}$) and standard deviation ($\sigma_{spot}$).  Since the spot's contribution to the overall CCF will be phase dependent from an observer's point of view, we modulate $A_{spot}$ in such a way that it has lesser effect on the wings of G$_{\rm star}$.

\begin{figure*}
    \centering
    \begin{tabular}{cc}
  \includegraphics[width=0.51\textwidth]{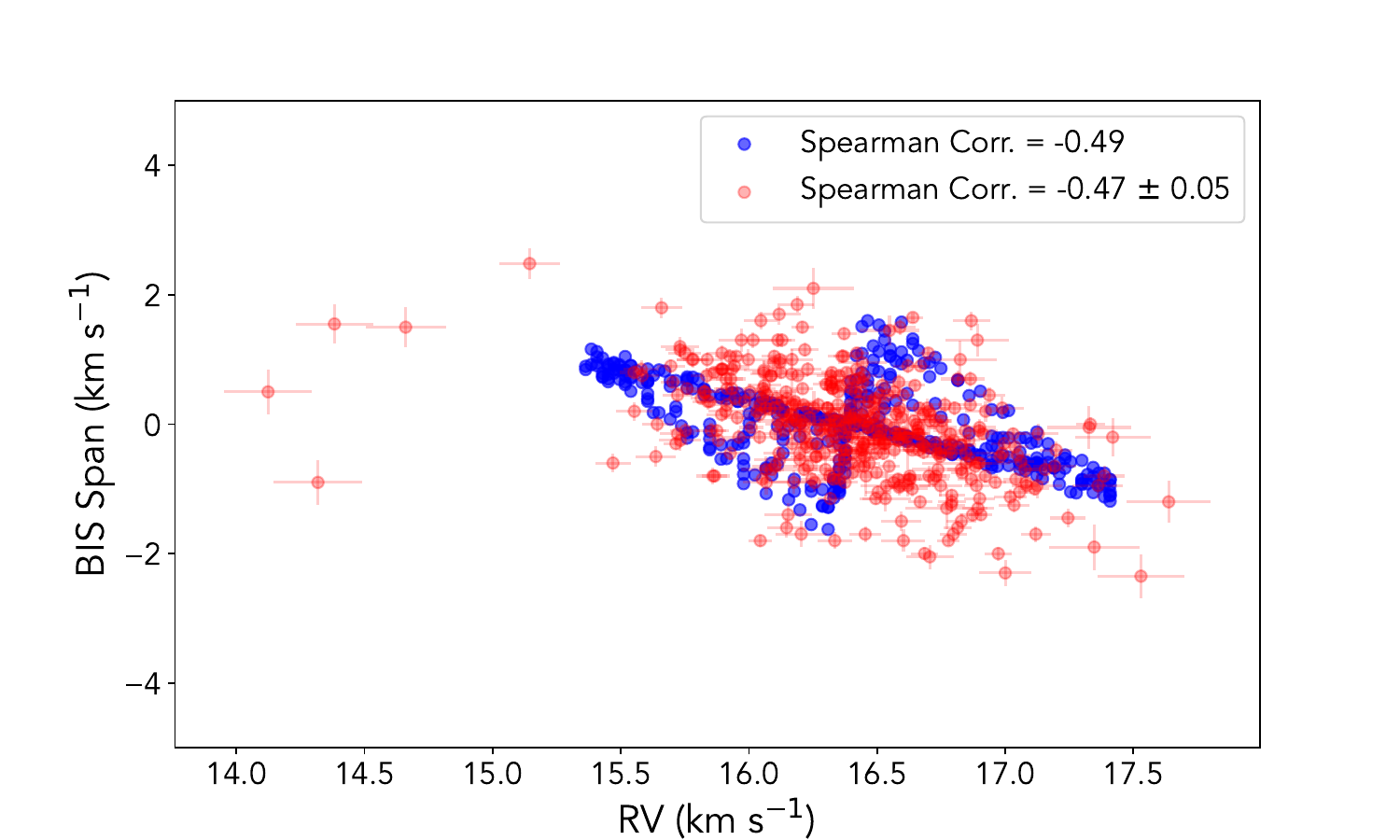}  
 \includegraphics[width=0.51\textwidth]{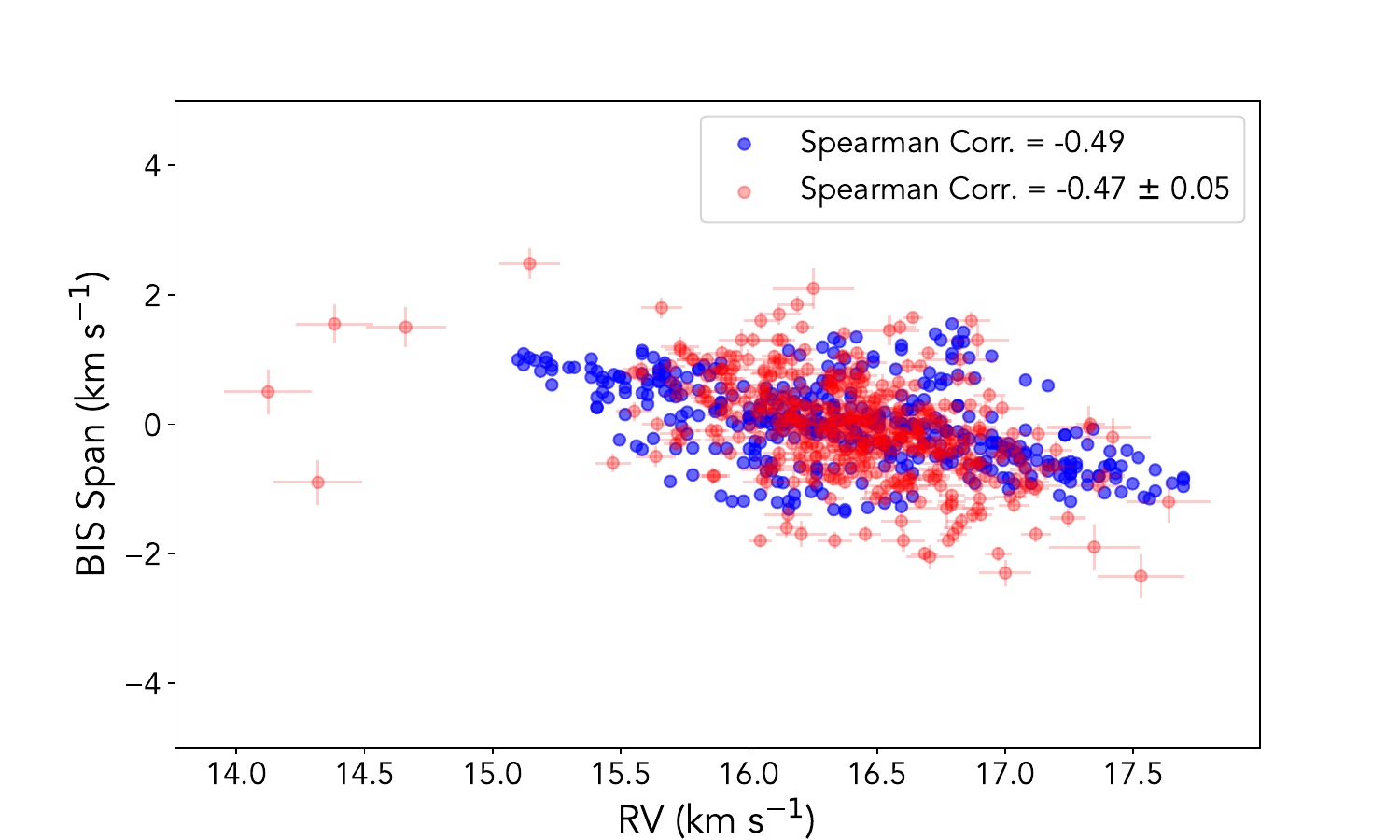}  
    \end{tabular}
    \caption{Plot of observed (red) and modelled (blue) RVs vs. BIS span for case 1 (left panel): G$_{\rm spot}$ modulated at 9 d and G$_{\rm star}$ kept fixed and case 2 (right panel): spot modulated at 9 d and star Doppler shifted with semi-amplitude of 0.3 km s$^{-1}$ at 25.2 d period. Individual point errorbars are also plotted.}
    \label{fig:bis_model}
\end{figure*}

In the first case, we sinusoidally modulate G$_{\rm spot}$ with a $\sim$ 9\,d period, keeping G$_{\rm star}$ stationary at a mean value of $\sim$ 16.39 km s$^{-1}$, corresponding to the observed mean radial velocity of CI Tau. The BIS span is computed from the combined CCF models (star + spot) and a Spearman coefficient of $\sim$ $-$0.49 is calculated from the model (Figure \ref{fig:bis_model}, left panel). In this case, since the star is not Doppler shifted, the only effect of the spot is to rotationally modulate the bisector lines about the mean velocity of the star. The RV vs. BIS span plot results in a distinctively ordered shape as can be seen \citep[similar to][]{hebrard2014}, and in this case we are unable to statistically reproduce the observed BIS span vs. RVs plot of CI Tau (as discussed below), represented by red circles in Figure \ref{fig:bis_model}.

In the second scenario, we sinusoidally modulate G$_{\rm spot}$ with a $\sim$ 9\,d period and we incorporate a periodic Doppler shift into G$_{\rm star}$, characterized by a 25.2-day period and a semi-amplitude of 0.3 km s$^{-1}$ as determined from our Keplerian orbital fit (see Section \ref{section:esp_spi_rvs}). A translational Doppler shift introduced into G$_{\rm star}$ due to Doppler reflex induced by the planet, induces a horizontal scatter in the modelled BIS span, which results in a better agreement in the overall shape of the observed BIS span (Figure \ref{fig:bis_model}, right panel). We experiment for different parameter values for the spot's Gaussian in our model and we find that a value of $\sim$ 0.7 for the ratio ($\sigma_{spot}$/$\sigma_{star}$) and $\sim$ 0.1 for the ratio $A_{spot}$/$A_{star}$, gives the best model match to the observed BIS span in this case. We also experiment with different initial phases for the periods with which G$_{\rm spot}$ and G$_{\rm star}$ are modulated. The results are independent of phases.

By employing the relationship between the radial velocity (RV) amplitude of the spot, its latitude ($l$), and the star's rotation ($\Delta V_{\text{rad}} = 2v\sin(i)\cos(l)$, as established by \citealt{bouvier2007} and \citealt{mcginnis2020}, and taking into account a value of $9.5 \pm 0.5$ for $v\sin(i)$ as reported by Donati et al. 2023 (submitted), our analysis suggests that the latitude of the cold spot should be close to zero, aligning with the best-fit solution observed in the RV vs. BIS span plot. This inference leads us to hypothesize that the observed spectral line-shape distortion attributed to cold spot activity may result from a combination of contributions, not only from the high-latitude spot \citep{donati2020}, but potentially from other equatorial cold spots as well, as suggested by Donati et al. 2023 (submitted).

Our model illustrates the impact of a translational shift in the star's CCF when accompanied by spot activity, a phenomenon previously examined by \citet{simola2019}. Additionally, as \citet{hebrard2014} demonstrated, the signature of spots exhibits a distinctive 'S-type' pattern. We can only reproduce this pattern in case 1, where the star remains fixed and only the spot modulates. However, the observed BIS span vs. RV plot exhibits significant scatter around the mean, almost resembling an ellipse, even if the respective errors are considered. We are able to replicate this behaviour to a considerable extent when the star's center of mass is in motion in tandem with the spot's motion. 

\begin{figure}
  \centering
     \includegraphics[width=9.5cm]{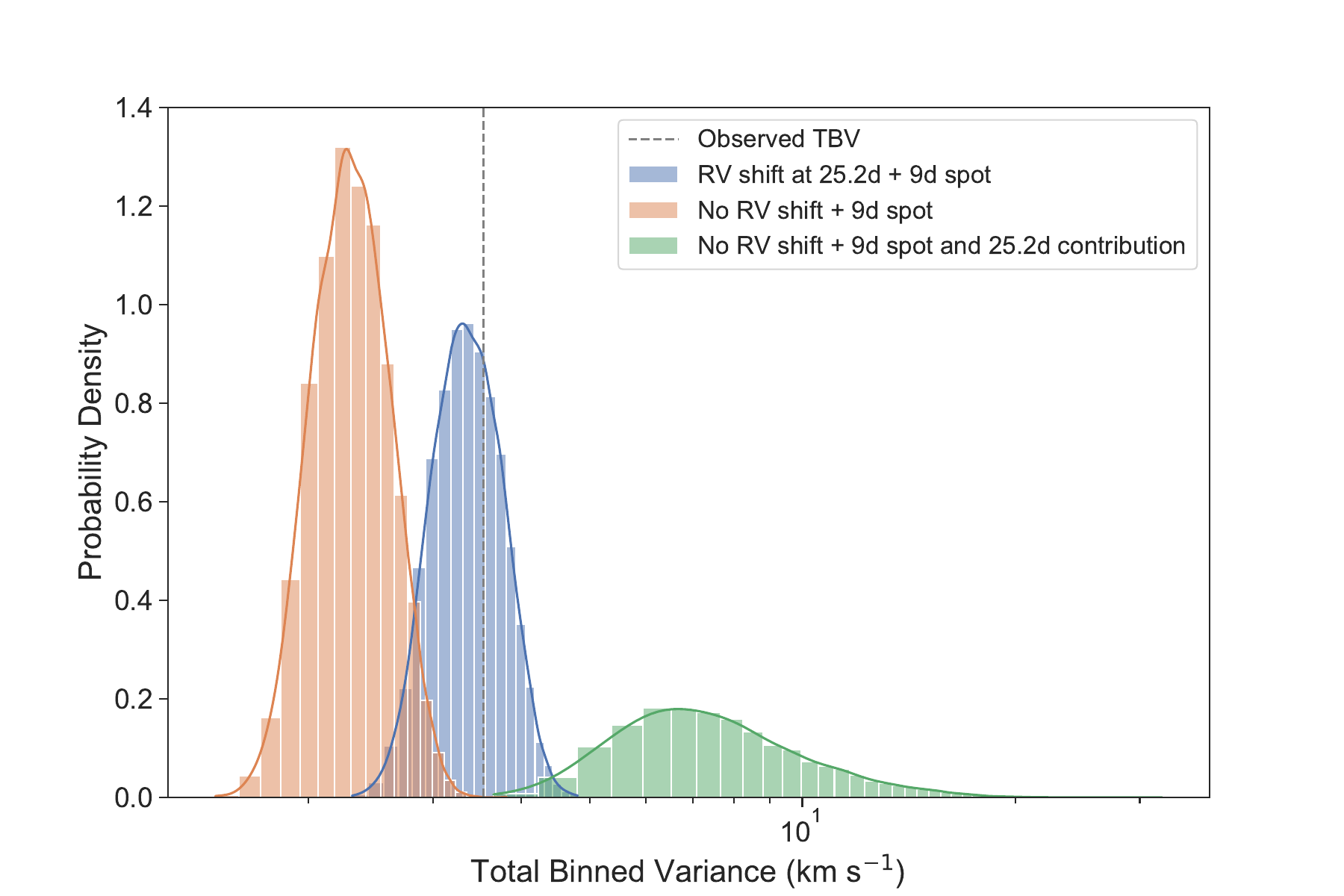}
      \caption{Plot of TBV vs. probability density after 10000 MC simulations for three cases: 1) when the star is at rest at spot modulated at 9 d and 2) when the star is Doppler shifted with a semi-amplitude of 0.3 km s$^{-1}$,  keeping all other parameters the same as in case 1. The orange and blue distributions represent TBVs of the BIS span plot for cases 1 and 2, respectively. Another scenario related to contribution from material from the inner disk on the CO lines, that could cause the 25.2-d periodicity, has also been tested and the distribution is shown in green. The grey line is the observed TBV for SPIRou BIS span vs. RVs plot.}
         \label{fig:tbv_pdf}
  \end{figure}

Since in both cases the spearman coefficients of the model and data are not very different from each other, although the shapes of the modelled BIS span vs. RV plot differs, we attempt to quantify the shape and the degree of scatter using a simple statistical method that computes a total binned variance (TBV) of the BIS span at binned RVs. For the observed data we compute a TBV 3.54 km s$^{-1}$. We carry out a Monte-Carlo simulation in which we consider uniformly distributed random values for the depth, mean and standard deviation of both the spot's and star's Gaussian profiles. For the latter, the values were uniformly distributed between the errors we obtained from the mean-profile fit of the SPIRou CCFs. We run 10,000 iterations in each case (case 1: fixed star and case 2: Doppler shifted star), generating probability density functions (PDFs, Figure \ref{fig:tbv_pdf}) of the TBV in each case. The results strongly indicate that the probability of obtaining the observed shape of the BIS span vs. RV plot solely through spot activity is exceedingly low ($\le$ 1\%). Our analysis suggests the observed BIS span vs. RV plot shape agrees best with a scenario in which we have combined effect of spot activity and Doppler reflex motion induced by a companion.

We also experiment on a scenario discussed by Donati et al. 2023 (submitted), which speculates that the 25.2 d period could be a result of a non-axisymmetric structure in the inner disk at around 0.16 au with a line profile contribution on the CO lines. This disc material with a Keplerian velocity of 57.4 km s$^{-1}$ could create a line profile contribution of around 3.8 times larger in FWHM in the CO photospheric lines compared to the observed LSD profiles and is expected to modulate RVs at a 25.2 d periodicity. Our veiling measurements unequivocally demonstrate that the CO lines have a photospheric origin. Specifically, when we subtract a template veiled to match the level of veiling observed in CI Tau from the original spectra, the residuals reach the noise level at the location of the CO lines. Assuming that there is still some undetected contribution from a non-axisymmetric material at 0.16 au, we test the scenario by adding this contribution to the total line profile (i.e., spot modulation at 9 d period + line profile contribution due to structure modulated at 25.2 d + fixed star). Our computed the TBV (shown in green distribution in Figure \ref{fig:tbv_pdf}) suggest this is statistically a less probable scenario to account for the observed BIS vs. RVs plot presented in Figure \ref{fig:bis_model}. 


\subsection{Combined SPIRou and ESPaDOnS RVs orbital analysis} \label{section:esp_spi_rvs}
\begin{figure*}
  \centering
    \includegraphics[width=18cm]{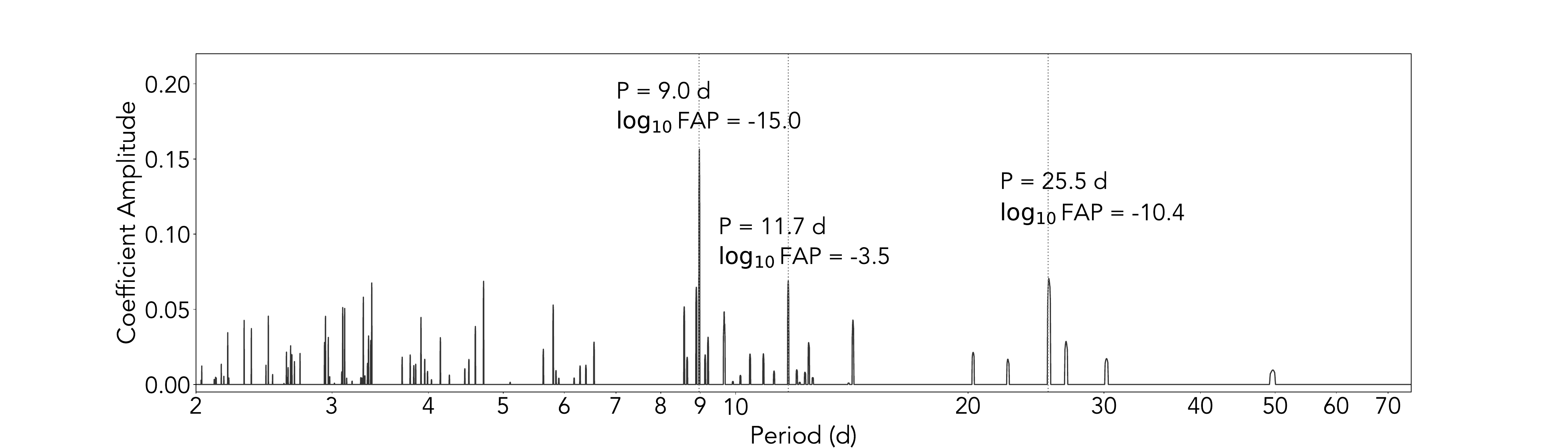} \\ 
     \vspace{0.1pt} 
    \includegraphics[width=18cm]{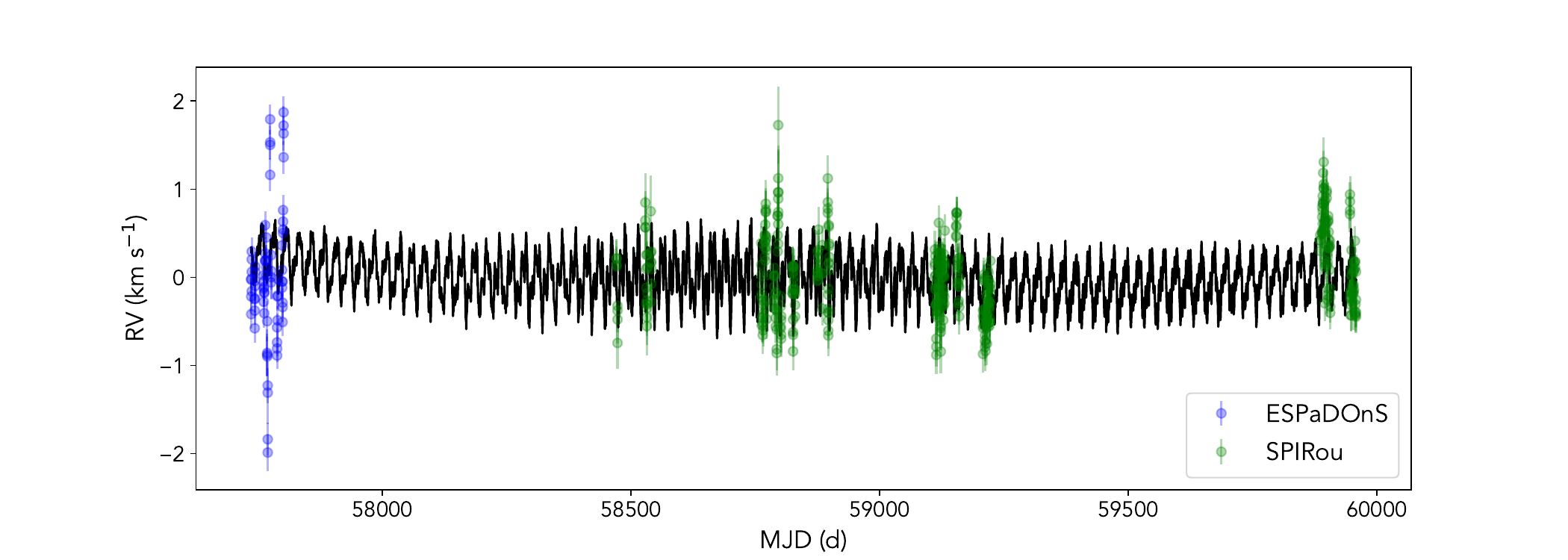} \\
    \includegraphics[width=10cm]{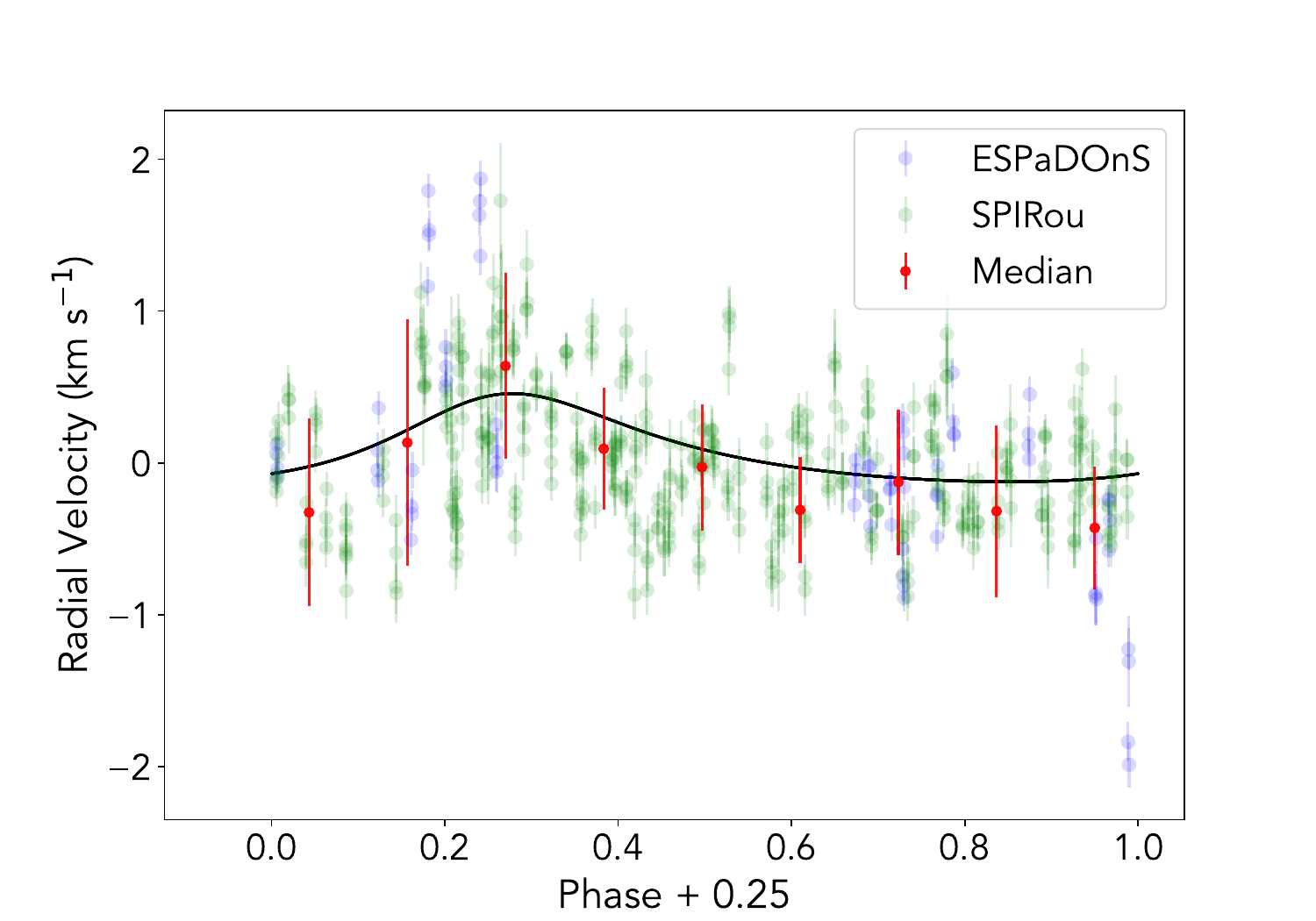} \\
    
      \caption{RV orbital analysis. Top: L1 periodogram of the full ESPaDOnS and SPIRou RVs. The 25.5\,d period peak, the spot modulation period peak at 9.0\,d and the 11.6\,d period peak, with their associated FAPs are marked.
      Middle panel: best-fit QP-GP combined model (9 d spot + 25.2 d planet) to the RVs. Bottom panel: best fit Keplerian orbit model (black curve) overplotted on RVs cleaned from the spot GP model. We introduce an arbitrary phase space shift of 0.25 to enhance the clarity of the orbital variation. The red dots illustrate a binned median of the data points, with bin sizes chosen to maintain uniform point distribution within each bin. The error bars on the red dots represent the standard deviation within each bin.}
         \label{fig:RVs_all_analysis}
  \end{figure*}

We carry out our RV orbital analysis on the combined ESPaDOnS and SPIRou RVs, spanning over $\sim$ 8 years. We show the L1 periodogram of this dataset in top panel of Figure \ref{fig:RVs_all_analysis}. Dominant periodicities are seen at 9.0 d, 11.7 d and 25.2 d\footnote{an additional long term period peak is seen at 444 d in the L1 periodogram.}.

We conducted a joint Markov Chain Monte Carlo (MCMC) fitting process for both the stellar activity spot and Keplerian orbital components using the combined RV data. The MCMC procedure was implemented using the Python package \textsc{emcee} and the full fitting process is described in Appendix \ref{section:mcmc_qp_gp_kepler}. We set uniform priors for all the QP-GP hyper-parameters:  GP amplitude ($h$), spot period ($P_{\text{spot}}$), smoothing parameter ($\lambda_{P_{\text{spot}}}$), and evolutionary time scale ($\lambda_{e_{\text{spot}}}$). The Keplerian orbit parameters include the orbital period ($P$), eccentricity ($e$), semi-amplitude ($K$), time at periastron passage ($T_0$), and the argument of periastron ($\omega$). We incorporate a zero semi-amplitude orbit model into our MCMC priors, specifically to accommodate cases where only a spot GP model is applicable. An observation close to middle of the data was chosen as initial guess for $T_0$. A Lucy and Sweeney test \citep{lucy1971} was performed on the residuals between a circular model and an eccentric model fit to the data. We find a $p$-value of 0.000851, which suggests an eccentric orbit is preferred over a circular one.

The priors and best fit parameters resulting from the MCMC fit are listed in Table \ref{tab:tab_kep_params}. In Figure  \ref{fig:RVs_all_analysis}, the best fit Keplerian model is overplotted in phase with the RV data cleaned from the spot GP model.

 We use the best-fit Keplerian parameters to determine the mass function given by:

\begin{equation}
f(m_1, m_2)={\frac {m_{2}^{3}\ \mathrm {sin} ^{3}i}{(m_{1}+m_{2})^{2}}}={\frac {P_{\mathrm {orb} }\ K^{3}}{2\pi G}}(1-e^{2})^{3/2}.
\end{equation}

Where $G$ is the gravitational constant. Since different values have been reported for the central star mass \citep[0.8 $\leq$ $m_1$ $\leq$ 0.9 M$_{\sun}$,][]{guilloteau2014, simon2019} and disk inclinations \citep[50$^{\circ}$ $\leq$ $i$ $\leq$ 70$^{\circ}$,][]{clarke2018,soulain2023}, we determine the range of possible values for the companion mass, assuming the orbital plane is aligned with one of the disks. The error on the companion mass was obtained considering the taking the entire inclination and mass ranges together with the errors obtained on the Keplerian orbital parameters listed in Table \ref{tab:tab_kep_params}. Our result is consistent with a $\sim$ 3.6 $\pm$ 0.3 M$_\mathrm{jup}$ planet in a highly eccentric orbit of $0.58_{-0.06}^{+0.05}$ eccentricity and at a semi-major axis of $\sim$ 0.17 $\pm$ 0.08 au from the central star.

\begin{table*}[htbp]
     \caption{Priors and best fit parameters for MCMC global fit of the Keplerian orbit and 9\,d spot. }
    \centering
   \renewcommand{\arraystretch}{1.3} 
    \begin{tabular}{l l l}
        \hline
        Parameter & Prior & Fit Value \\
        \hline
          \rule{0pt}{10pt} 

         Keplerian orbital parameters & & \\
         \rule{0pt}{10pt}
        $P_{\text{orb}}$ [d] & $\mathcal{U}[21, 29]$ & $25.2_{-1.8}^{+1.7}$ \\
        $\rm ecc.$ & $\mathcal{U}$[0.0, 0.99] & $0.58_{-0.06}^{+0.05}$  \\
        $K$ [km s$^{-1}$] & $\mathcal{U}$[0.0, 1.0] & $0.30_{-0.03}^{+0.02}$ \\
        $T_0$ [d] & $\mathcal{U}$[58780, 58880] & $58821.4_{-2.4}^{+2.1}$ \\
        $\omega$ [deg] & $\mathcal{U}$[0, 360] & $344_{-19}^{+10}$ \\
       
        \hdashline 
        \rule{0pt}{2pt}
        QP-GP hyper-parameters & & \\
         \rule{0pt}{10pt} 
        GP amplitude ($h$) [km s$^{-1}$] & $\mathcal{U}$[0.0, 1.8] & $0.34_{-0.04}^{+0.03}$ \\
        Spot period ($P_{\text{spot}}$) [d] & $\mathcal{U}$[8.0, 10.0] & $9.0_{-0.5}^{+0.4}$ \\
        Smoothing ($\lambda_{\text{p}}$) & $\mathcal{U}$[0.1, 3.0] & $0.8_{-0.07}^{+0.09}$\\
        Evol. time scale ($\lambda_{\text{e}}$) [d] & $\mathcal{U}$[50, 1500] & $446_{-30}^{+33}$ \\
        White noise amplitude ($\sigma_{RV}$) [km s$^{-1}$]  & $\mathcal{U}$[0, 1] & $0.06_{-0.007}^{+0.005}$  \\
               \hdashline 
        Fit report & & \\
               RMS data [km s$^{-1}$] & - & 0.50 \\
               RMS residuals [km s$^{-1}$] & - & 0.32 \\
               Red. $\chi^2$ fit & - & 7.0 \\
        \hline
    \end{tabular}
    \label{tab:tab_kep_params}
\end{table*}

\subsection{Planet RV detection limit in spot activity}

We conducted experiments to simulate the detection limit of the periodogram peak corresponding to an injected planet with a 25.2-day orbital period, in the presence of varying levels of spot activity. To achieve that we utilised Gaussians constructed for the star (G$_{\rm star}$) and spot (G$_{\rm spot}$) described in Section \ref{sec:spirou_bis}.

\begin{figure}
  \centering
    \includegraphics[width=9.5cm]{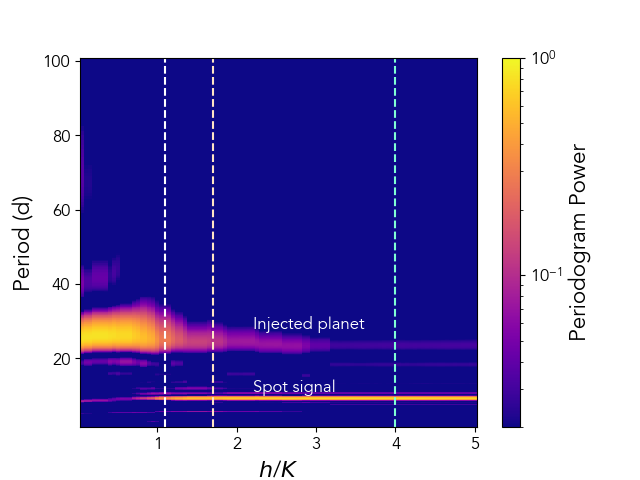} 
      \caption{2d periodogram of simulated RVs to show the peak detection limits of an injected planet at 25.2\,d period in the presence of spot modulation at 9\,d. The value of $h/K$ we derive for CI Tau is shown with the white dashed line at 1.1. The aquamarine line shows the detection limit in the optical where we expect the activity RV signals to be $\sim$ 4 times larger. The light orange dashed line at 1.7, depicts $h/K$ we expect in the H-band.}
         \label{fig:spot_to_star_amplitude_ratio_2d}
  \end{figure}

To mimic the Doppler shift caused by the injected planet onto the star, we introduced a periodic shift of 25.2 days on \textbf{G$_{\rm star}$}. This shift was maintained constant at a semi-amplitude value of $K$ = 0.3\,km s$^{-1}$, which was determined through a Keplerian orbital fit (Section \ref{section:esp_spi_rvs}). We then modulate G$_{\rm spot}$ with a period of 9 d and an amplitude ($h$). At each modulation step, we compute a total profile (G$_{\rm star}$ + G$_{\rm spot}$) and record the modulation steps as time. We fit a Gaussian to the total profile each time to extract the RV (as we would do for a normal observed CCF). The procedure was executed for various values of $h$, yielding a 1D periodogram for each $h/K$. The combination of these individual results produces a 2D periodogram, as illustrated in Figure \ref{fig:spot_to_star_amplitude_ratio_2d}. Subsequently, we determine the threshold at which the peak corresponding to the planet can be consistently detected in the periodogram. Various initial phases between the spot and the injected planet were investigated during this exploration.

Our simulation results indicate that a clear detection of the period peak associated with the injected planet occurs when the ratio $h/K$ is approximately 1.0 or less (see Figure \ref{fig:spot_to_star_amplitude_ratio_2d}). The planet's signal is only marginally discernible for $h/K$ values between $\sim$\,1.0 and 2.0 and almost not detected for $h/K$ values higher than $\sim$\,2.0. For CI Tau, our analysis yields an $h$ value of 0.34\,km s$^{-1}$, determined through GP regression of the spot model in the RVs and $K$=0.3\,km\,s$^{-1}$. This leads to an $h$/$K$ ratio of $\sim$\,1.1, where the RV signal arising from spot rotation slightly predominates over the Doppler signal from the injected planet. Nevertheless, the planet's signal can still be marginally detected (indicated by the white dashed line in Figure \ref{fig:spot_to_star_amplitude_ratio_2d}). This provides a potential explanation for the limited detection of the 24\,d peak in the SPIRou RVs (Figure \ref{fig:gls_rvs_compa}) that were derived from spectral lines spanning the entire SPIRou spectral range, encompassing bluer wavelengths that are particularly susceptible to the influence of spot activity.

In addition, as shown by \cite{miyakawa2021}, activity levels exhibit chromatic behaviour and can affect optical lines up to a level of $\sim$\,4 times more than in the red. If we consider spot RV signal to be 0.34\,km\,s$^{-1}$ from our GP fit to CO RVs and 4 times more in the optical, i.e. an $h/K$ value of $\sim$\,4 and will be most likely not detected in lines most affected by spot activity. Donati et al. 2024 (submitted) demonstrate a approximately 1.5-fold increase in activity levels from the K-band (where the CO line are) to the H-band, within the SPIRou domain. Considering this factor, we anticipate $h/K$ to be 1.7 in the H-band. As depicted in Figure \ref{fig:spot_to_star_amplitude_ratio_2d}, this is likely why the planet's RV signal will be hardly detected in the H-band.

Moreover, in the radial velocities obtained from ESPaDOnS, a comparable situation is observed. As depicted in Figure \ref{fig:bis}, the BIS span values span a range corresponding to an amplitude of 2.5 km\,s$^{-1}$, encompassing both spot activity and the planet's signal. Taking half of that value implies an $h/K$ of 4.2, possibly explaining the limited detection of the planet's RV signal in the ESPaDOnS data.

Interestingly, our 2D periodogram reveals faint appearance of period peaks at approximately 11.6 days, 40 days and longer periods exceeding approximately 60 days. These peaks become more noticeable when $h$/$K$ values are below approximately 1.0, where the injected planet's and spot's signals are comparable. We suspect that these peaks could possibly be beat periods arising from the combined influence of the 24\,d, 9\,d peaks and other secondary beat peaks that might arise from any of these periods.

We emphasise that our model solely illustrates the collective impact of both a spot and an introduced planet on the overall CCF and BIS span. One of our key assumptions is that the spot's influence on the entire CCF profile can be approximated by a Gaussian distribution. However, in reality, it can possess a more complex structure, as demonstrated in \citet{donati2020}. Furthermore, our model does not account for other quasi-periodic or stochastic effects stemming from magnetic fields or other factors related to stellar activity or accretion, which could potentially introduce additional features in the BIS span and CCF RVs and which could make the planet's signal detection even harder at bluer wavelengths.

\section{Discussion} \label{section:discussion}
\subsection{Periods in light curve} \label{section:discussion_photometric_periods}
Our detailed analysis of the K2 and LCOGT light curves show periodicities at 6.6\,d, 9\,d (spot), 11.5\,d, 14.2\,d and 24\,d/26.4\,d. We examine whether this 6.6 d period can be understood through an unstable accretion state for CI Tau. Using 3D numerical simulations, \citet{blinova2016} have shown that the accretion state (stable or unstable) can be determined by computing the ratio of the truncation radius to the corotation radius, $r_{\rm t} / R_{\rm cor}$. They estimated that accretion is unstable if $r_{\rm t} / R_{\rm cor}$ $\lesssim$ 0.7. Using literature values \citep[e.g.][]{donati2020} for the mass-accretion rate, dipolar magnetic field strength, stellar mass and radius of CI Tau, we find that $r_{\rm t} / R_{\rm cor}$ $\approx$ 0.3-0.57 \footnote{In order to obtain the disk truncation radius, we use the scaling from \citet{pantolmos2020}.}. This value of $r_{\rm t} / R_{\rm cor}$ indicates that accretion in CI Tau is unstable. As shown in \citet{blinova2016}, unstable accretion states can produce significant periodogram peaks at values lower than the stellar period. In addition, such peaks were observed, through wavelet analysis, not to be stable over the entire duration of the simulations (see Figures 3 and 5 of \citealt{blinova2016}). Hence, we hypothesise that the 6.6 d period identified in both the K2 light curve and in some parts of the RVs, could be attributed to the influence of this specific accretion regime.

A slightly longer period than the spot modulation is seen to be coherent at 11.5\,d and 11.6\,d in the K2 and LCOGT light curves, respectively. The exact origin of this period is still not clearly understood, but there are clues that it could potentially arise due beating effects between the 6.6\,d and 14.2\,d peaks. In the same way, the 14.2\,d period peak could well be a beat period between the 9\,d and 24\,d peaks. One other mechanism giving rise to these peaks could be time-variable accretion. In CTTS, the mass-accretion rate can vary by up to a factor 0.4\,dex on time scales of days to weeks \citep[e.g.][]{costigan2014,zsidi2022}. Furthermore, recent studies \citep[e.g.][]{venuti2021} have demonstrated that non-steady accretion onto CTTS is expected to produce periodic signals in light curves at periods larger than the stellar period. This particular CTTS feature is further supported by numerical simulations of star-disk interactions  \citep[e.g.][]{romanova2005,zanni2013}, which have shown that magnetospheric accretion is not a steady-state process and can occur in quasi-periodic cycles with timescales longer that the stellar period. In summary, periods shorter than the stable long period of 24\,d, can be related to magnetospheric accretion processes, that are hard to interpret.

The only stable period that seem to be consistently detected in both the K2 and LCOGT light curves and with high statistical significance, is the longer period at 24\,$\pm$\,5 d in the K2 data and at 26.4\,$\pm$\,0.6\,d in the LCOGT data, both of which are consistent within their respective errors. \citet{teyssandier2020} presented hydrodynamical simulations of hot Jupiters orbiting protoplanetary discs, exploring how a giant planet's mass, orbital separation, and eccentricity regulate the accretion rate at the inner disc edge. They found that a massive, eccentric planet can drive pulsed accretion at the inner edge of the disc, modulated by the planet’s orbital frequency. The amplitude of accretion variability generally increases with the planet mass and eccentricity in their model. They specifically studied CI Tau, showing that the observed luminosity variability in the star can be explained by accretion signatures modulated by an eccentric planet  at 9 d. However, since it has been well established that the 9 d is due to spot modulation and not a planet, it is equally possible that an eccentric planet with a longer period modulates the accretion onto the star. This aligns well with our results, which indicate an eccentric and massive planet is indeed present in CI Tau at an orbital period of 25.2\,d.

\begin{figure}
  \centering
    \includegraphics[width=9.5cm]{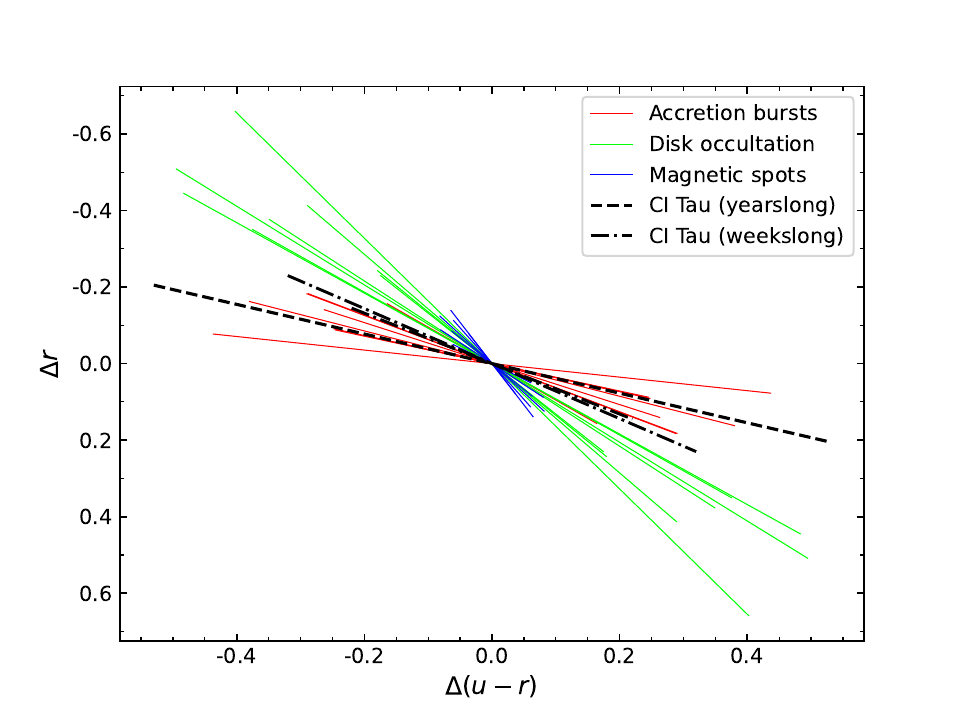} 
      \caption{Optical variability amplitudes and colour slopes of CI Tau, following the classification of \citet{venuti2015}. It suggests CI Tau's colour behaviour is consistent with accretion-dominated sources.}
         \label{fig:color_color_venuti}
  \end{figure}

An alternate mechanism that could account for the 24\,d/26.4\,d periodicity observed in the light curves may result from the fact that CI Tau is observed at a relatively high inclination, approximately 70 degrees.This may lead to variations in brightness due to circumstellar matter periodically obstructing the star's photosphere (resulting in variable extinction events). This occulting material might be situated in the innermost region of the accretion disk, at a distance of $\sim$ 0.16\,au from the central star. We note, however, that distinct colour behaviours are expected for a CTTS with variability dominated by circumstellar extinction as opposed to accretion dynamics, and the specific colour properties of CI Tau clearly favour the latter scenario. To test this point, we followed the approach of \citet{venuti2015}, who explored the typical correlation trends between optical variability amplitudes and simultaneous blue-filter colour variability for TTSs dominated respectively by magnetic spots, accretion features, and recurrent extinction by circumstellar material. We used our LCOGT time series photometry, as well as archival U-to-R monitoring data \citep[e.g.][]{herbst1999}, to derive colour slopes for CI Tau over days to years timescales and compare them with the range of typical behaviours described in \citet{venuti2015} for different types of TTS variables (see, in particular, their Fig. 7). This analysis revealed that CI Tau displays a colour behaviour consistent with the properties of accretion-dominated sources (and inconsistent with being extinction-dominated) on all timescales from a few weeks (comparable to the rotational timescales) to several years, as shown in Figure \ref{fig:color_color_venuti}.

\subsection{Periodic variability at 25.2\,d seen in the spectroscopic RVs} \label{sec:discussion_rvs}
Our bisector analysis shows that in the presence of solely the spot-induced 9 d modulation period, the CCFs will be modulated periodically in a way so that only a rotational effect on the bisector will be detected about the star's mean RV. By introducing a periodic component of 25.2\,d as a Doppler shift in the CCFs, with an RV semi-amplitude obtained from our Keplerian model, we notice that the observed bisector span is more accurately replicated.

We test detection limits of a 25.2\,d planet in CI Tau by varying the amplitude of the spot signal while keeping the Doppler signal due to the injected planet constant. We demonstrate that when the Doppler shift caused by the planet is on the same order of magnitude as the RV amplitude of the spot, the planetary signal can be detected to a significant statistical level in the periodogram of RVs that are less influenced by spot activity.
In addition, we note that RV Doppler shift induced by the planet will be of the same RV amplitude in spectral lines at all wavelengths, but the shift will be diluted at bluer wavelengths which are known to be most affected by the star's activity.

We reduce the effect of spot activity in our RV analysis by using photospheric lines in the SPIRou wavelength domain as red as possible, since the spot contrast is known to decrease at longer wavelengths \citep[e.g.][]{miyakawa2021}. We then combine all RV data from ESPaDOnS and SPIRou to perform a global MCMC of the spot activity and the Keplerian orbit. Our comprehensive spectroscopic analysis, statistically supports the notion that the star exhibits a Doppler reflex motion with a 25.2 d period due to a massive and eccentric planet. We do not entirely dismiss other potential, although less probable mechanism(s). These mechanisms are considered less likely since they should induce a Doppler reflex in the spectral lines, a point we substantiate through our comprehensive bisector analysis.

\subsection{Eccentric planet}
Our best-fit Keplerian RV model indicates an eccentric planet with an eccentricity of 0.58$^{+0.05}_{-0.06}$. Given the highly eccentric massive planet located close to the actively accreting star, we expect a pulsed-accretion mechanism operating in CI Tau as proposed by \citet{teyssandier2020}. Their models predict only an eccentric massive planet in CI Tau can produce a periodic variability in the accretion which will be imprinted on the star's light curve, making it periodically brighter at phases when the planet is at its closest approach to the star. With the detection of the same coherent periodicity in both K2 and LCOGT light curves, we suspect a pulsed-accretion could be a likely mechanism to create such a modulation in the light curves.

\begin{figure*}
  \centering
      \includegraphics[width=18cm]{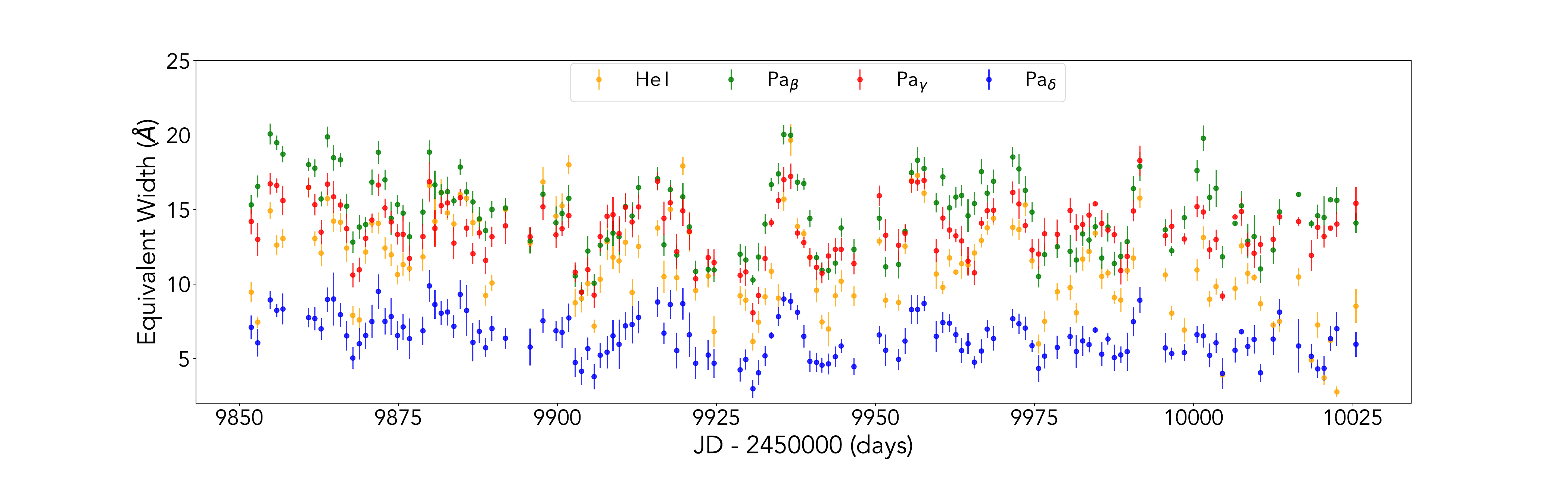} 
    \includegraphics[width=18cm]{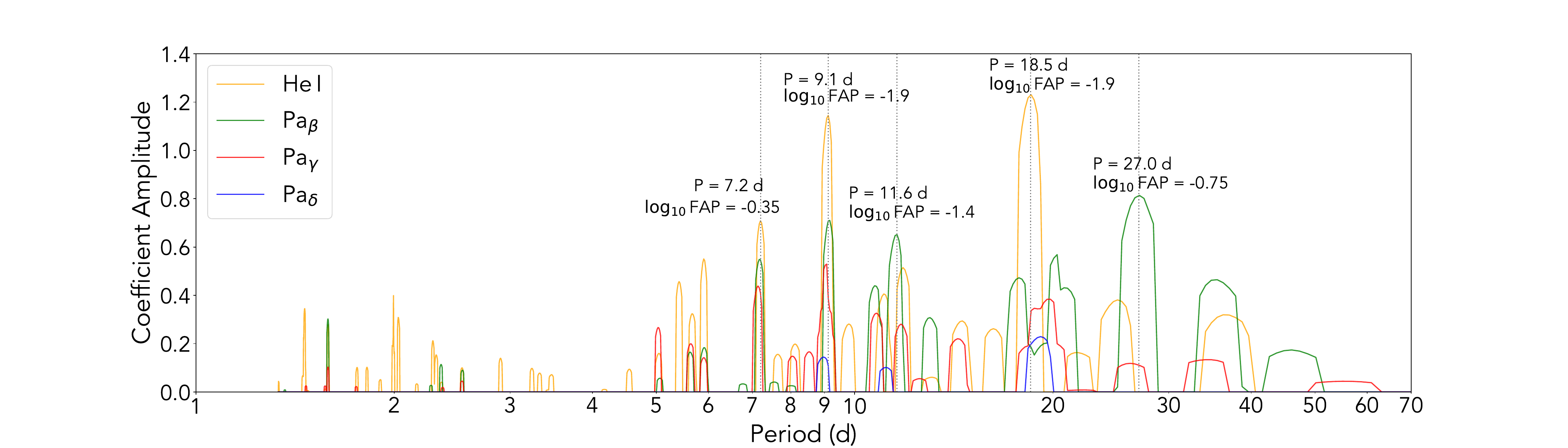} 
    \includegraphics[width=12cm]{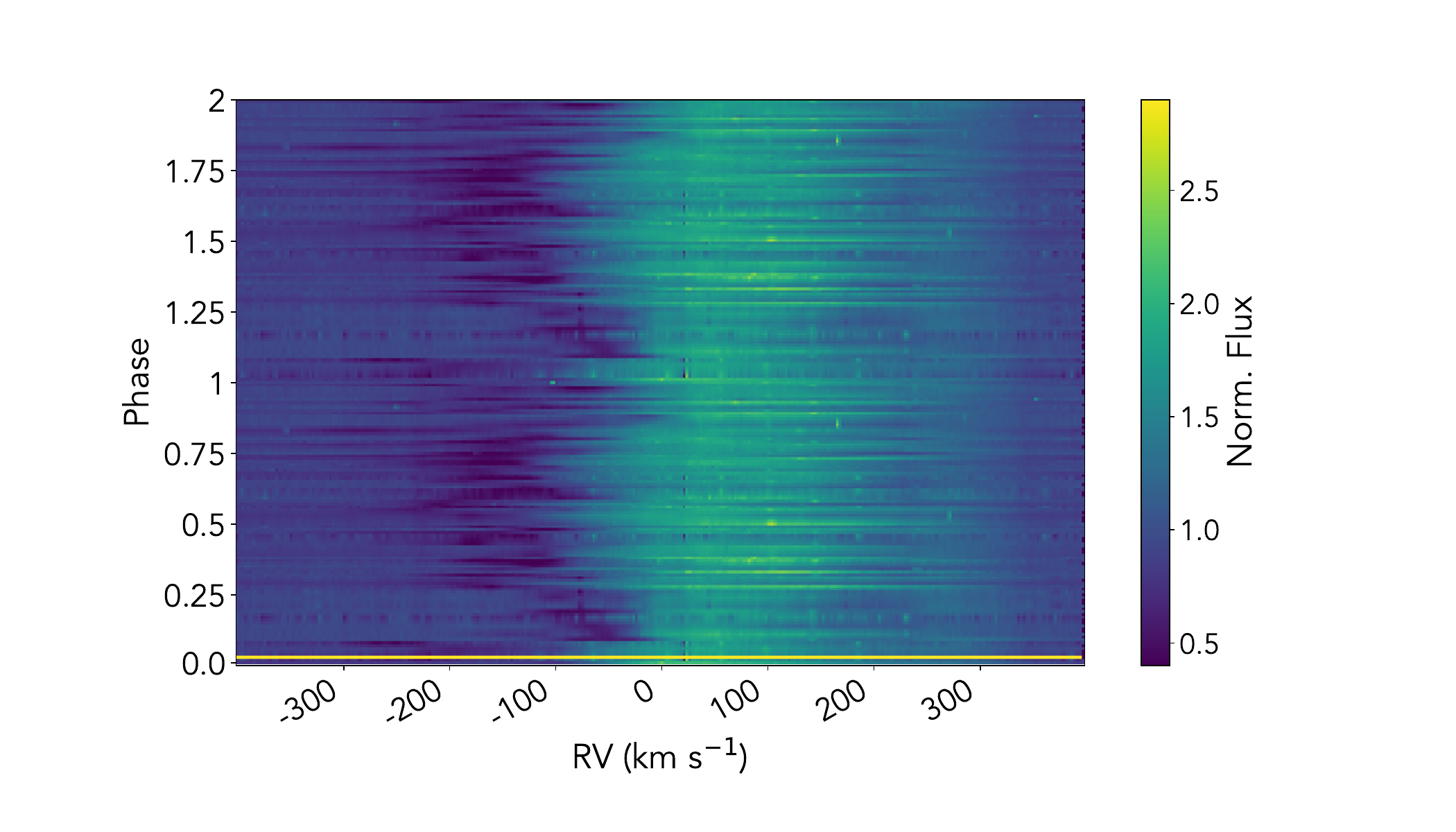} 
      \caption{Top: Time series of EW measured for He\,I, Pa$\beta$, Pa$\gamma$ and Pa$\delta$ lines. Middle: L1 periodogram of EW. A clear peak is detected in the strongest line tracing accretion (Pa$\beta$) at $\sim$ 27.0 d. Additional peaks that are relatively significant are seen at 7.2 d, 9.1 d (Spot), 11.6 d, 18.5 d (see text). Bottom: Trailed plot of the He\,I line phased at a period of 19.5 d showing high velocity outflows traced by the blue absorption component in the line. The colour bar represents the normalised flux.}
         \label{fig:extra_ew}
  \end{figure*}

To test this, we analyse the equivalent width (EW) of the Pa$\beta$ line in the ExTrA data, which is indicative of accretion-induced activity. A distinct peak at approximately 27.0 d is detected (see Figure \ref{fig:extra_ew}). We note that the periodicity peak is highest in Pa$\beta$ EW periodogram, likely because it is the strongest line among the three Paschen lines (Pa$\beta$, Pa$\gamma$ and Pa$\delta$). The detection of this periodicity in this line, potentially provides observational support for the planet-induced phase-dependent pulsed accretion model proposed by \citet{teyssandier2020}.

An additional peak at around 18.5 d is detected in the EW of the HeI-10830 line (Figure \ref{fig:extra_ew}). This period closely aligns with a periodicity of approximately 19.5 d observed in the blue-shifted absorption component of the HeI-10830 line (see bottom panel of Figure \ref{fig:extra_ew}). We posit that this may be tracing periodic outflows with velocities ranging from $\sim$ 30 - 200 km s$^{-1}$ within the system. Outflows such as stellar winds and magnetospheric ejections have shown similar velocities in numerical simulations of star-disk interactions \citep{romanova2009,pantolmos2020, bouvier2023} 

Furthermore, the system exhibits a significant excess in the near-infrared (NIR) portion of its spectral energy distribution (SED). \citet{muley2021} conducted an SED analysis of CI Tau, concluding the presence of an inner dust-free cavity. This observation is further corroborated by recent interferometric imaging of the very inner region of the CI Tau disk \citep{soulain2023}, revealing a resolved inner disk cavity with a radius of $21\pm2~R_{\star}$ (0.18$\pm$0.02 au). This discovery aligns with the findings of \citet{muley2021} and suggests that a close-in planet might have played a role in carving this inner gap. Recent investigations into the inner CO disk of CI Tau by \citet{kozdon2023} indicate a CO disk break occurring at approximately $0.14\pm0.08~\rm au$, in close agreement with the expected planet location. Hydrodynamical simulations performed by \citet{debras2021} demonstrate that a Jupiter-mass planet, after migrating into a low-density gas cavity, can acquire a high eccentricity of $0.3-0.4$. Furthermore, \citet{baruteau2021} propose that a $\sim$ 2 Jupiter mass planet can achieve an eccentricity of $0.25$ within a few million years (Myr), resulting in pronounced asymmetries in the gas surface density within the cavity.

Since CI Tau is only a 2\,Myr old object, we do not expect an inner planet to have circularised in this lifetime. Recent models by \citet{terquem2021}, aim at assessing the time scale needed for circularisation due to tides raised by a Jupiter-sized planet around a solar mass star. The authors find that tidal dissipation in a Jupiter-sized planet leads to a circularisation time scale of 1\,Gyr for an orbital period of 3 days and this relation stays flat after orbital periods of 3 d. Their value is somehow higher than what was predicted in the models by \citet{rodriguez2010}. The authors used numerical simulations to investigate the long-term tidal evolution of a single-planet system. In their model, they included variations in the semi-major axis and eccentricity of the relative orbit, taking into account the influence of both planetary and stellar tides across various planet types. They determine the critical value of eccentricity at which the stellar tide surpasses the planetary tide in a Sun-Jupiter type system and they computed a circularisation time scale of at least 0.1\,Gyr, which is around 2 orders of magnitude higher than CI Tau's age. 

\section{Conclusions} \label{section:conclusion}

Our comprehensive examination of the CTTS CI Tau has unveiled a statistically significant periodicity of approximately 25.2\,d within the system. Through an extensive analysis of bisector span, we have gathered evidence that this 25.2\,d period is related to a Doppler shift in the RVs least affected by the effects of spot activity. Our analysis supports the existence of an eccentric planet with a mass of 3.6\,$\pm$\,0.3 M$_\mathrm{jup}$ in orbit around CI Tau at a semi-major axis of 0.17\,$\pm$0.08 au. We investigated alternative scenarios that could potentially induce such a periodicity in the system but were unable to identify one. Furthermore, we have demonstrated that in a actively accreting system like CI Tau, spot activity signals might attenuate the Doppler reflex signal of a planet in RVs associated with bluer wavelengths, which would make the planetary signal hard to detect in that domain.

CI Tau's photometric light curves from the K2 mission and LCOGT display other significant periods that range between $\sim$ 6.6 d and $\sim$ 47 d. The only period shorter than the 9 d spot rotation period is found to be a 6.6 d period, which is seen only in the K2 light curve and not in the LCOGT light curve that is taken $\sim$ 3 to 6 years later. This period is also detected in the ESPaDOnS RVs, but is not very clear in the SPIRou RVs. We hypothesize that this period is not stable over the long term and may be related to unstable accretion on the stellar surface. Other periods that are slightly longer than the spot rotation period are 11.5 d and 14.2 d, which we think could be related to either beating effects or non-steady accretion variability on the stellar surface, although we cannot reach any clear conclusions about these periodicities.

Our study outlines the difficulty of detecting inner-embedded planet(s) in the $\sim$ inner 0.1 au's of young CTTS systems which are dominated by activity. We demonstrate that when working with light curves and RVs , one of the main limitations to the detection of exoplanets in active stars like CI Tau, is the different sources of variability induced by the accreting star itself, rather than the instrument's RV precision. Only planets that induce Doppler signals in the RVs that are of the same order or higher than the spot activity signals in highly accreting systems like CI Tau, are likely to be detected in lines that are least affected by the star's activity. Our study indicates that, unless future research convincingly establishes an alternative explanation for the 25.2\,d periodicity, our findings, in conjunction with recent independent studies of the inner regions of CI Tau, strongly support the planet hypothesis as the most plausible explanation. Importantly, this marks the first identification of an inner embedded planet in a CTTS, highlighting that even in systems with pronounced activity signals, the detection of inner embedded planets is achievable.

\begin{appendix}

\section{CCF profiles of CO lines from SPIRou observations and their mean profile.}

Each SPIRou observation consists of four subexposures to measure Stokes V, taken at different orientations
of the polarimeter and used to compute the non-polarized and the circularly polarized profiles. We treat each sub-exposure as an individual spectrum and compute a mean CCF of photospheric lines in the CO region for each of them.

\begin{figure}
  \centering
     \includegraphics[width=9cm]{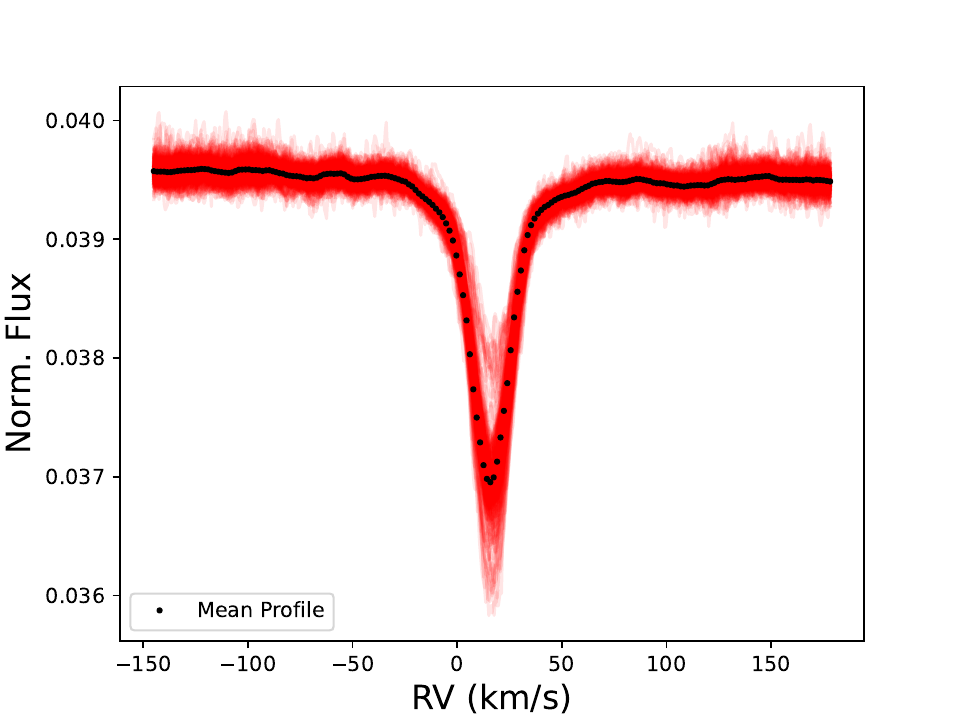}
      \caption{Plot of all CCF mean profiles of the photospheric CO lines (in red) and their mean profile shown in black.}
         \label{fig:spirou_ccf_profiles}
  \end{figure}

In Figure \ref{fig:spirou_ccf_profiles}, we show the average of all the CCF profiles in black dotted points. Note that barycentric velocity correction and local normalisation to the continuum level were applied to all the spectra, achieved through a polynomial function fitting the continuum.

\section{MCMC sampling plot for best-fit QP-GP hyper-parameters and RV orbital parameters. } \label{section:mcmc_qp_gp_kepler}

We perform an MCMC sampling to simultaneously fit a combined QP-GP model for the spot and a Keplerian orbit model. Our analysis begins by defining a log-likelihood function which quantifies the agreement between the combined model and observed data through a chi-squared fit. Subsequently, we computed the posterior probability distribution, which integrates information from both the log-prior and log-likelihood functions. To explore parameter space, we employed a total of 256 walkers and executed 10,000 iterations during the MCMC sampling process. The initial positions for these walkers were set to correspond to initial parameter guesses. 

\begin{figure*}
  \centering
    \includegraphics[width=12cm]{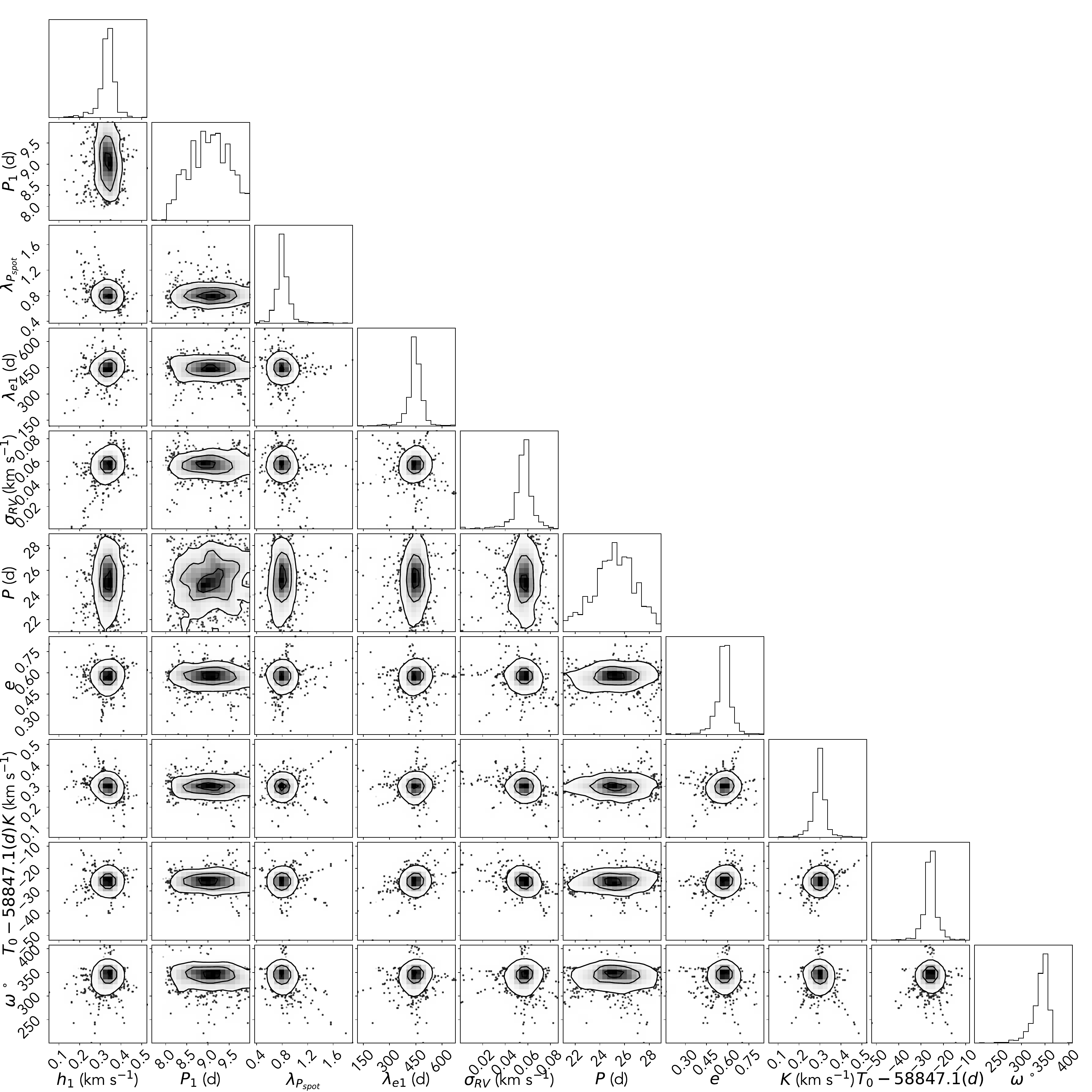}
      \caption{Corner plot of QP-GP hyper-parameters and best-fit Keplerian orbit model parameters to the combined ESPaDOnS and SPIRou RV dataset, after 10,000 MCMC iterations with 256 walkers thinned to 256,000 points per sampled parameter.}
         \label{fig:corner_plot_mcmc}
  \end{figure*}

We set uniform priors for all the QP-GP hyper-parameters and Keplerian orbit parameters with values listed in Table \ref{tab:tab_kep_params}. The resulting MCMC chain was recorded, and posterior distributions were generated based on the entire 10,000 iterations with 256 walkers. For improved sampling efficiency and to reduce autocorrelation within the chain, we applied a thinning factor of 10. Consequently, this led to a final distribution of 256,000 points for each sampled parameter (Figure \ref{fig:corner_plot_mcmc}).

\end{appendix}

\begin{acknowledgements}
Rajeev Manick (RM) would like to respectfully dedicate this study to his dearly departed mother (Uma Devi Manick), who sadly passed away around the time of detection. The author kindly requests that any forthcoming studies referring to the planet be named "\textit{Uma}" in her honour. We thank Jean-François Donati for discussions and comments about the results presented in this paper. We express our gratitude to the anonymous referee for providing valuable suggestions that significantly contributed to the improvement of the paper. We acknowledge support from the European Research Council (ERC) under the European Union’s Horizon 2020 research and innovation programme (grant agreement No 742095; SPIDI: Star-Planets-Inner Disk-Interactions, http://www.spidi-eu.org). We thank Dr. Evelyne Alecian, Dr. Oscar Barragan, Dr. Pia Cortes-Zuleta, Dr. Anthony Soulain and Dr. Catherine Dougados for fruitful discussions on the RV analysis and Hugo Nowacki for providing the code for creating the 2d periodogram of the ESPaDOnS RVs. This research made use of NASA’s Astrophysics Data System; SciKit GStat (M. M\"{a}licke et al. 2021), SciPy (Virtanen et al. 2020b); NumPy (Harris et al. 2020); matplotlib (Hunter 2007); and Astropy, a community-developed core Python package for Astronomy (Astropy Collaboration et al. 2018). We thank the director of LCOGT for a generous allocation of discretionary time. E.M. acknowledges funding from FAPEMIG under project number APQ-02493-22 and a research productivity grant number 309829/2022-4 awarded by CNPq, Brazil. This research was made possible through the use of the AAVSO Photometric All-Sky Survey (APASS), funded by the Robert Martin Ayers Sciences Fund and NSF AST-1412587. RM acknowledges routines used from the IvS repository of KU Leuven to create trailed plot shown in Figure 16.  
\end{acknowledgements}

%
%

  \bibliographystyle{aa} 
  \bibliography{refs.bib} 

\end{document}